\documentclass[twocolumn]{aastex63}
\usepackage{xspace}
\usepackage{changepage}
\usepackage{rotating}
\usepackage{multirow}
\usepackage{comment}

\defcitealias{Castignani2021}{Paper I}

\received{June 15, 2021}
\accepted{October 20, 2021}

\newcommand{\ha}{H$\alpha$\xspace}

\shorttitle{Virgo Filaments II: the catalog}
\shortauthors{VF collaboration}
\begin{document}
\title{Virgo Filaments II:  Catalog and First Results on the Effect of Filaments on galaxy properties}

\author[0000-0001-6831-0687]{Gianluca Castignani}
\affiliation{Dipartimento di Fisica e Astronomia, Alma Mater Studiorum Università di Bologna, Via Gobetti 93/2, I-40129 Bologna, Italy}
\affiliation{INAF - Osservatorio  di  Astrofisica  e  Scienza  dello  Spazio  di  Bologna,  via  Gobetti  93/3,  I-40129,  Bologna,  Italy}
\affiliation{Institute of Physics, Laboratory of Astrophysics, Ecole Polytechnique Fédérale de Lausanne (EPFL), Observatoire de Sauverny, CH-1290 Versoix, Switzerland}
\author[0000-0003-0980-1499]{Benedetta Vulcani}
\affiliation{INAF- Osservatorio astronomico di Padova, Vicolo Osservatorio 5, I-35122 Padova, Italy}
\author[0000-0001-8518-4862]{Rose A Finn}
\affiliation{Department of Physics and Astronomy, Siena College, 515 Loudon Road, Loudonville, NY 12211, USA}
\author{Francoise Combes}
\affiliation{Observatoire  de  Paris,  LERMA,  Collège  de  France,  CNRS,  PSL  University,  Sorbonne  University,  75014, Paris}
\author[0000-0002-9655-1063]{Pascale Jablonka}
\affiliation{Institute of Physics, Laboratory of Astrophysics, Ecole Polytechnique Fédérale de Lausanne (EPFL), Observatoire de Sauverny, CH-1290 Versoix, Switzerland}
\affiliation{GEPI, Observatoire de Paris, Universit\'{e} PSL, CNRS, Place Jules Janssen, F-92190 Meudon, France}
\author[0000-0001-5851-1856]{Gregory  Rudnick}
\affiliation{University of Kansas, Department of Physics and Astronomy, 1251 Wescoe Hall Drive, Room 1082, Lawrence, KS 66049, USA}
\author[0000-0002-5177-727X]{Dennis Zaritsky}
\affiliation{Steward Observatory, University of Arizona, 933 North Cherry Avenue, Tucson, AZ 85721-0065, USA}
\author{Kelly Whalen}
\affiliation{Department of Physics \& Astronomy, Dartmouth College, 6127 Wilder Laboratory, Hanover, NH 03755, USA}
\author{Kim Conger}
\affiliation{University of Kansas, Department of Physics and Astronomy, 1251 Wescoe Hall Drive, Room 1082, Lawrence, KS 66049, USA}
\author{Gabriella De Lucia}
\affiliation{INAF - Astronomical Observatory of Trieste, via G.B. Tiepolo 11, I-34143 Trieste, Italy}
\author{Vandana Desai}
\affiliation{Spitzer Science Center, California Institute of Technology, MS 220-6, Pasadena, CA 91125, USA}
\author{Rebecca A Koopmann}
\affiliation{Department of Physics \& Astronomy, Union College, Schenectady, NY, 12308, USA}
\author{John Moustakas}
\affiliation{Department of Physics and Astronomy, Siena College, 515 Loudon Road, Loudonville, NY 12211, USA}
\author{Dara J Norman}
\affiliation{National Optical Astronomy Observatory, 950N Cherry Avenue, Tucson, AZ 85750}
\author{Mindy Townsend}
\affiliation{University of Kansas, Department of Physics and Astronomy, 1251 Wescoe Hall Drive, Room 1082, Lawrence, KS 66049, USA}
\correspondingauthor{GC: gianluca.castignani@unibo.it}

\begin{abstract}
\noindent Virgo is the nearest galaxy cluster; it is thus ideal for studies of galaxy evolution in dense environments in the local Universe. It is embedded in a complex filamentary network of galaxies and groups, which represents the skeleton of the large scale Laniakea supercluster. Here we assemble a comprehensive catalog of galaxies extending up to $\sim12$ virial radii in projection from  Virgo to revisit the Cosmic Web structure around it. 
This work is the foundation of a series of papers that will investigate the multi-wavelength properties of galaxies in the Cosmic Web around Virgo.
We match spectroscopically confirmed sources from several databases and surveys including  HyperLeda, NASA Sloan Atlas, NED, and ALFALFA. The sample consists of $\sim7000$ galaxies. 
By exploiting a tomographic approach, we identify 13 filaments, spanning several Mpc in length. Long  $>17~h^{-1}$~Mpc filaments, tend to be thin ($<1~h^{-1}$~Mpc in radius) and with a low density contrast ($<5$), while shorter filaments show a larger scatter in their structural properties. Overall, we find that filaments are a transitioning environment between the field and cluster in terms of local densities, galaxy morphologies, and fraction of barred galaxies. Denser filaments have a higher fraction of early type galaxies, suggesting that the morphology-density relation is already in place in the filaments, before galaxies fall into the cluster itself.  We release the full catalog of galaxies around Virgo and their associated properties. \end{abstract}


\section{Introduction} \label{sec:intro}
Galaxies in the Universe are not distributed uniformly at the megaparsec scales. Large galaxy redshift surveys have revealed that the Universe has a prominent web-like structure made by dense clusters and groups, elongated filaments, planar sheets and voids, called the Cosmic Web \citep{Tifft_Gregory1976,Joeveer1978, Bond1996}. 
Galaxies are continuously funneled into higher density cluster environments through  filaments, which host $\sim$40\% of the galaxies \citep[e.g.,][]{Jasche2010, Tempel2014_f, Cautun2014}. 
Therefore, the analysis of filamentary structures can carry insights on the assembly history of large scale structures.

Characterizing the Cosmic Web and flow of galaxies in the nearby Universe is not an easy task and many strategies have been proposed, based on either observations \citep{Tully2013,Tully2016} or simulations \citep[e.g.,][]{Libeskind2018,Libeskind2020}. These methods often rely on the study of the geometry of the galaxy density field or of the tidal field to reconstruct the Cosmic Web, which indeed consists of a set of structures that are anisotropic in shape (e.g. elongated filaments), multi-scale (groups, clusters, and filaments that can extend from a few to 100 Mpc), and are intricately connected \citep[see, e.g.,][]{Cautun2014}. 
The absence of both a common definition for the cosmic filaments and a unique operative procedure to identify the filamentary structures, as well as the lack of fields observed with a very high sampling rate, have been major obstacles in investigating not only the structure of the Cosmic Web, but also its impact on galaxy evolution.

Despite difficulties, filamentary structures of the Cosmic Web have been identified in both simulations \citep[e.g.,][]{Aragon-Calvo2010, Cautun2014, Chen2015, Laigle2018, Kraljic2019,Kuchner2020,Kuchner2021,Rost2021b}
and galaxy surveys \citep[e.g.,][]{Tempel2014_f, Alpaslan2014, Chen2016, Laigle2018, Kraljic2018, Malavasi2017,Malavasi2020b, Malavasi2020a}. 
Many works have also suggested that filaments affect the evolution of the integrated properties of galaxies \citep[e.g.,][]{Koyama2011, Geach2011, Sobral2011, Mahajan2012, Tempel_Libeskind2013, Tempel2013, Zhang2013,  Pintos2013, Koyama2014, Santos2014, Malavasi2017,  Mahajan2018} and the distribution of satellites around galaxies \citep{Guo2014}, at any redshift, but results are still controversial. Overall, filament galaxies tend to be more massive, redder, more gas poor and have earlier morphologies than galaxies in voids \citep{Rojas2004, Hoyle2005, Kreckel2011, Beygu2017, Kuutma2017}. Some studies have also reported an increased fraction of star-forming galaxies  
\citep{Fadda2008, Biviano2011, Darvish2014, Porter2007, Porter2008,  Mahajan2012}, and higher metallicities and lower electron densities \citep{Darvish2015} in filaments with respect to field environments.

Other studies even found evidence of a distinct impact of filaments on galaxy properties and different gas phases. \cite{Vulcani2019_fil} showed that ionized H$\alpha$ clouds in some filament galaxies extend far beyond what is seen for other non-cluster galaxies. The authors suggest this may be due to effective cooling of the dense star forming regions in filament galaxies, which ultimately increases the spatial extent of the H$\alpha$ emission.  Even atomic HI gas reservoirs are impacted by the filament environments  \citep{Kleiner2017, Odekon2018, BlueBird2020,Lee2021}. The global properties of galaxies' gas reservoirs as a function of distance to the filament and local density are still debated. Some studies claimed that galaxy and halo properties (e.g., luminosities, masses, accretion rate, concentration) depend mostly on local density, while the filament environment has no additional effects beyond the ones related to the local density enhancement \citep{Yan2013,Eardley2015,Brouwer2016,Goh2019}. 

Further investigations are therefore clearly needed. Our approach is to focus on the area around Virgo, the benchmark cluster in the local Universe. It is embedded in a complex filamentary network as it indeed belongs to the Laniakea supercluster \citep{Tully2014}. The closeness of Virgo and its associated high spectroscopic completeness makes its field ideal for studies of galaxy evolution over a large range in  environments.

Numerous studies have  characterized the galaxy population of the Virgo cluster \citep{Kim2014}, and evaluated the associated atomic and molecular gas content \citep{Giovanelli2005, Chung2009, Boselli2014a,Boselli2014c, Boselli2014b}, dust \citep{Davies2010}  stellar masses \citep{Ferrarese2012}, and star formation \citep{Boselli2014_galex}. However, galaxies in the surrounding regions have received relatively little attention. 
\cite{Tully1982} identified prolate and oblate overdensities of galaxies connected to the cluster. Nonetheless, due to the limited size of their sample, these elongated structures were not clearly revealed as conventional narrow filaments. A better characterization of these structures requires improved statistics from larger galaxy samples, particularly those with fainter galaxies. Building upon Tully's results, \cite{Kim2016} used the seventh release of the Sloan Digital Sky Survey \citep[SDSS,][]{Abazajian2009} combined with the HyperLeda catalog \citep{Makarov2014} to more firmly identify the filamentary structures within an extensive volume around the Virgo cluster. 
While providing a detailed characterization of the filaments around the Virgo cluster, \cite{Kim2016} did not release their environmental classification.

In  \citet[][from now on Paper I]{Castignani2021} we therefore assemble an independent catalog of Virgo and the surrounding volume.  We accomplish this by matching and vetting several existing catalogs, with the intent of releasing a comprehensive catalog of galaxies in the filaments around Virgo, extending out to $\sim12$ virial radii in projection (i.e., $\sim24$~Mpc) from the cluster, with a small fraction ($7\%$) of  galaxies reaching even higher distances, up $\sim40$~Mpc from Virgo.
Two strengths of this catalog are that (1) we have high completeness because we merge sources from multiple catalogs of local galaxies, and (2) we have low contamination because we visually inspect every source in our sample.

Here we describe more in detail our adopted procedure and release the catalog (see Appendix~\ref{sec:catalogs}). With respect to \citetalias{Castignani2021}, we refine both the source catalog, by visually inspecting each object to remove duplicates, stars and ``shredded" galaxies and by excluding galaxies in the southern hemisphere where SDSS has poor spectroscopic coverage, and the filament definition (see Sec.\ref{sec:fil}).

Overall, we consider a larger survey area in the Nor then emisphere  than that covered by \cite{Kim2016}. This allows us to identify and characterize additional filamentary structures to the North and East of the Virgo cluster that were not identified by \cite{Kim2016}.

This catalog forms the foundation of a series of papers aimed at investigating the effect of the filament environment on processing the gas of galaxies, and on global properties such as star formation and stellar content. 
The first exploration of the catalog has been presented in  \citetalias{Castignani2021} where we analysed spatially integrated CO and HI observations for a subset of filament galaxies. We found a clear progression as one moves from field to filament and cluster in that galaxies in denser environments have lower star formation rate, a higher fraction of galaxies in the quenching phase, an increasing proportion of early-type galaxies, and a decreasing gas content. In addition, galaxies in the densest regions in filaments tend to be deficient in their molecular gas reservoirs, which fuel star formation.  These results suggested that processes that lead to star formation quenching are already at play in filaments. 
Following this study we are  carrying out  follow-ups at different wavelengths, with the aim of  linking the galaxy stellar properties to the galaxy gas content.  In particular, we will investigate the physical mechanisms responsible for the pre-processing using ongoing high resolution observations in both CO and HI. In parallel, for a few hundred filament galaxies, we are conducting an \ha  imaging survey to map the spatial distribution of the hot gas and to derive integrated star-formation rates. All these campaigns will be described in forthcoming papers.

The outline of this paper is the following. In Sect.~\ref{sec:parent_catalog} we describe how we build the catalog of galaxies around Virgo. In Sect.~\ref{sec:characterize_environment} and \ref{sec:other_envs} we characterize the Cosmic Web environment around Virgo. 
In Sect.~\ref{sec:compare_envs} we contrast the different parameterizations of environment and in Sect.~\ref{sec:prop_galaxies} we investigate the interplay between galaxy properties and their cosmic-web environment and describe our results. In Sect.~\ref{sec:conclusions} we draw our conclusions and summarize the paper.

Throughout this paper, we assume a Hubble constant of $H_0 = 100~h~ {\rm km}^{-1}~{\rm Mpc}^{-1}$, where $h=0.74$ \citep[e.g.,][]{Tully2008,Riess2019}. Magnitudes are reported in the AB system. 

\section{The Spectroscopic Parent Catalog}\label{sec:parent_catalog}

To assemble a spectroscopic sample of galaxies around Virgo ($RA=187.70^\circ$, $DEC=12.34^\circ$, J2000), we start by creating a catalog from the union of HyperLeda \citep{Makarov2014}\footnote{http://leda.univ-lyon1.fr/}, the NASA Sloan Atlas\footnote{http://nsatlas.org} \citep{Blanton2011}, and the ALFALFA $\alpha100$ sample \citep{Haynes2018} in the region covered by 
$ 100^\circ < RA < 280^\circ$, $-1.3^\circ < DEC < 75^\circ$, and recession velocities $500 < v_r < 3300$~km/s. 
The southern limit coincides with the southern limit of the SDSS spectroscopic survey. We adopt this cut because we want high spectroscopic sampling to robustly identify and characterize filaments.  However, this choice is different from what was done by \cite{Kim2016} who also characterized Virgo filaments to the south.
The lower velocity cut is dictated by the need to avoid stars and galactic contamination.  The higher velocity cut is set by the need to include all filaments, which are mostly localed farther than Virgo \citep[$cz\sim1000$~km/s,][]{Mei2007}. 

To build the sample, we start with all sources from HyperLeda that are classified as galaxies. 
We then match the HyperLeda sources to version 1 of the NSA, using a search radius of 10\arcsec\ and maximum velocity offset of 300 km/s. This updated version of the NSA extends to larger distances and contains additional fitted parameters\footnote{https://www.sdss.org/dr13/manga/manga-target-selection/nsa/} relative to version 0 presented in \citet{Blanton2011}.
 
We initially allow for the same NSA source to be matched to multiple HyperLeda sources, and we later eliminate these duplicates by visual inspection (see below). We then append as new catalog entries any additional NSA sources that were not matched to HyperLeda. We repeat a similar match to version 0 of the NSA \citep{Blanton2011} because some of the sources and redshifts differ between the two versions of the NSA catalogs.  Versions 0 and 1 of NSA are complementary in terms of the number of galaxies that fall in the region of interest, which motivates the choice of querying both catalogs. 

We then match the list of HyperLeda+NSA galaxies to the ALFALFA $\alpha$100 sample \citep{Haynes2018}, limited 
to ALFALFA galaxies with $500 < v_r < 3300$~km/s.  The northern limit of the ALFALFA survey is ${\rm DEC} = 36^\circ$, so we do not have ALFALFA coverage for our full survey area.  However, 54\% of our $DEC < 36^\circ$ sources are matched to an ALFALFA source, and this provides a rich sample for future studies on how the atomic gas reservoir is affected by the filament environments.  As a blind HI survey, ALFALFA detects a higher fraction of low-mass star-forming galaxies relative to optical surveys \citep[e.g.][]{Durbala2020}.  However, at the relatively close distance of Virgo, we find only 9 ALFALFA sources that are not already in either the Hyperleda or NSA catalogs.   We add these 9 sources to our catalog.

We assign a position (RA, DEC) to each galaxy based on information in the source catalogs.  We assign HyperLeda coordinates if they are available.  If HyperLeda is not available, we then use NSA version 0, followed by NSA version 1, and ALFALFA.  We assign recession velocities by the same process.

Next, we add to the sample { 110} galaxies that have redshift-independent distances
in the NASA/IPAC Extragalactic Database compendium of distances based on primary and secondary indicators \citep[NED-D,][]{Steer2017}. These { 110} galaxies have redshift independent distances that correspond to cosmological velocities in the range of 500-3300~km/s, but they are missing in our catalog as their observed recession velocities are less than $<500$km/s. Some of these sources are Virgo cluster members which are located near the caustics and thus have the largest deviation in velocity with respect to that of Virgo. 

Finally, to compile as clean a sample as possible in the area of interest, we visually review each galaxy in our catalog to remove shredded galaxies, duplicates, and spurious objects. We also flag galaxies with nearby stars that might affect the photometry, and we recenter the coordinates of some galaxies, as needed. 

{ To identify any remaining stars, we cross-match with the star catalog used by the Legacy Survey\footnote{\url{https://portal.nersc.gov/cfs/cosmo/data/legacysurvey/dr9/masking/gaia-mask-dr9.fits.gz}}.  This catalog is built from Tycho-2 (MAG$_{\rm VT}<13$) and Gaia-DR2 sources (G$< 16$). We look for matches within $r < 10$\arcsec of our sources, and we find an additional 2 stars which we remove.}

While we require all of the sources to have a galaxy classification, we find that a number of HyperLeda sources are instead globular clusters in nearby galaxies, as identified by \cite{Ko2017}. We therefore remove all sources with prefix ``S'' in the \cite{Ko2017} catalog.  On the basis of our cleaning procedure mainly aimed at removing duplicates and shredded objets, we found that $\sim4\%$ of HyperLeda sources in our region of interest are misclassified as galaxies.

For each galaxy, we also query the NED server to get its official NED name.  We use the object name from HyperLeda as input if it is available.  If not, we then use the NSA name and the ALFALFA/AGC name.  If NED does not return a match by name for any of the catalog names, we then match the source by position, using a search radius of 10\arcsec.
We include the input name used in the NED search as well as the official NED name in our table.  Note that for some galaxies, we are not able to find a corresponding NED name. 

Our final sample contains { 6780} galaxies.  The contributions from the different input catalogs are broken down in Table~\ref{tab:parent}. The NSA v1 (v0) catalog provides { 157} (122) galaxies that are not in the v0 (v1) version of the catalog. We stress that only 9 galaxies are in the ALFALFA $\alpha100$ sample, but not in the union of the HyperLeda and NSA source samples. 

\begin{table}
\caption{Statistics of the parent sample \label{tab:parent}}
\centering
\begin{tabular}{|l|c|c|}
\toprule
catalog & \# of galaxies & fraction \\ 
\hline
final & { 6780} & 1 \\
HL & { 6622} & 0.98 \\
NSA v1 & { 5280} & 0.78 \\
NSA v0 & { 5245} & 0.77 \\
$\alpha$100 & 2336 & 0.34 \\
NED-D & 1959 & 0.29 \\
\hline
\end{tabular}
\end{table}

\begin{figure*}
    \centering
    \includegraphics[scale=0.4]{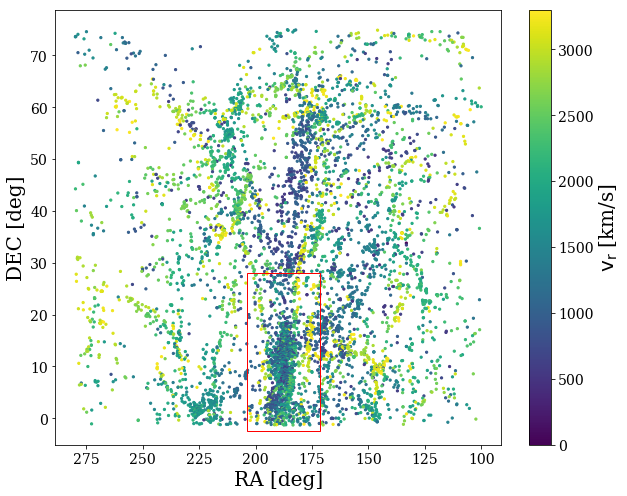}
    \includegraphics[scale=0.4]{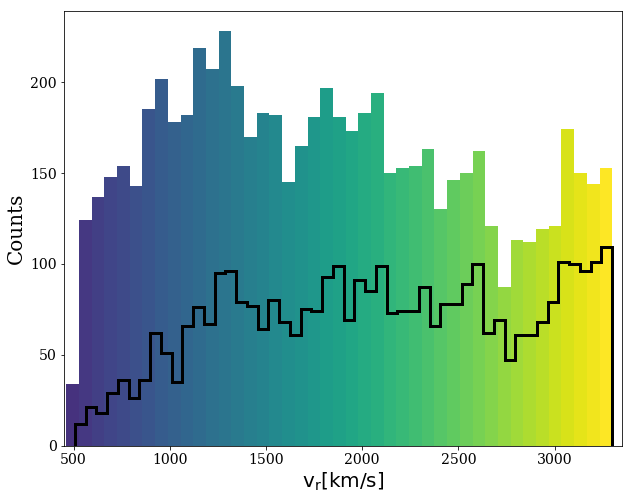}
    \caption{Left: Spatial distribution of galaxies around the Virgo galaxy cluster, up to $\sim$12 virial radii from its center, in projection. Points show galaxies coded according to their recession velocities. The red rectangle shows the region of the Virgo cluster, which is examined in more detail in Fig.\ref{fig:phase_space}. Right: Radial velocity distribution of all galaxies in our catalog.  Black solid line shows the distribution of galaxies brighter than the absolute magnitude completeness limit of the catalog ($M_r=-15.7$).     \label{fig:positions_vr}}
\end{figure*}

\subsection{Photometry}
We cross-match our catalog to the ninth public data release of the DESI Legacy Imaging Surveys (DR9, \citealt{Dey2019}), using a search radius of 10\arcsec. The Legacy Survey covers 14,000 square degrees of extragalactic sky visible from the northern hemisphere in three optical bands (g,r,z) and four infrared bands.  In this paper, we utilize only the r-band photometry from the DR9 catalogs to apply a magnitude cut and analyze galaxy properties in a absolute magnitude complete sample. 

 The available DR9 photometric catalogs are based on the Tractor fitting \citep{Lang2016}; like all automated photometry code, Tractor struggles with providing meaningful models to clumpy, well-resolved  galaxies.  We therefore have efforts underway to measure custom photometry from the Legacy imaging that is optimized for large, nearby galaxies.  In a forthcoming paper
 we will present the multi-band photometry for our entire catalog of sources in the field of Virgo, with a careful treatment of the extended galaxies ($>0.5$~arcmin in size). 

As most of the spectroscopic redshifts for galaxies in the catalog come from the SDSS, we adopt the SDSS completeness limit of $r=17.77$. This corresponds to an absolute limit of $M_r=-15.7$ at a distance modulus of 33.5, approximately the upper limit of the survey.

\subsection{The final catalog}\label{sec:final_catalog}
To summarize, we have assembled a catalog of galaxies with $500 < v_r < 3300$~km/s located in the region surrounding the Virgo cluster (up to $\sim12$ virial radii, i.e. 24~Mpc, in projection from the center of Virgo) by combining the sources present in HyperLeda, NSA (v0 and v1), ALFALFA, and NED-D. This catalog is cleaned from spurious sources, stars, and duplicates and represents a unique starting point to define the Cosmic Web around the Virgo cluster, as detailed in what follows. 

The final catalog (Table~\ref{tab:parent}) contains { 6780} galaxies, { 3528} of which above the absolute magnitude limit $M_r=-15.7$ \citep[$\simeq M^\star_r+3$,][]{Blanton2005b}.  
The subsample of galaxies with $M_r<-15.7$ corresponds to a volume limited sample, with the $M_r$ limit corresponding to the SDSS $m_r=17.77$ spectroscopic completeness limit at the maximum distance of the galaxies in our catalog.  As we will describe in Sect.~\ref{sec:fil}, we define different volume-limited subsamples appropriate for the distance range of each filament.

Figure \ref{fig:positions_vr} shows the projected  spatial distribution of the final sample, color coded by recession velocity (left panel), and the distribution of the recession velocity (right panel), for both the entire sample and sub-sample above the absolute magnitude limit. 

Hereafter, to identify galaxies in different global environments (Sect.~\ref{sec:characterize_environment} and Sect.~\ref{sec:other_envs_gr}), we will make use of the full catalog of $\sim7000$ galaxies. When we compute local densities to characterize the properties of galaxies in the different environments (from Sect.~\ref{sec:ld} onward) we will adopt the magnitude complete sample.

\section{The Virgo Cluster and its infalling filaments}\label{sec:characterize_environment}
In this Section we provide a characterization of the Cosmic Web around Virgo using the catalog of galaxies assembled above. We will rely on a widely used description of the cosmic flow around Virgo \citep{Mould2000} and on redshift independent distances, when available \citep{Steer2017}. 

To properly investigate the effect of the Mpc-scale environment on galaxy properties, it is necessary to provide both local and more global
parameterizations of the density as, depending on the scale probed, different physical processes might shape galaxy properties.  For example, the frequency of galaxy-galaxy interactions depends on the local density of galaxies, whereas gas accretion onto galaxies varies depending on whether the galaxy is a central or satellite galaxy in the parent halo mass.
After computing the distances for all galaxies (Sect.~\ref{sec:dist}), we will thus assign to all galaxies a global environment depending on whether they are in the Virgo cluster (Sect.~\ref{sec:cl}) or in filamentary structures (Sect.~\ref{sec:fil}). In the next Sect.~\ref{sec:other_envs_gr} we will further investigate the presence of groups within filaments and assemble a sample of pure field galaxies, aided by the \cite{Kourkchi2017} group catalog. We will finally evaluate local densities for all galaxies (Sect.~\ref{sec:ld}), regardless of all of the above memberships. 

We will then use these characterizations to i) determine the filament profiles (Sect.~\ref{sec:density_profiles}), ii) compare the different definitions of environment (Sect.~\ref{sec:compare_envs}), and iii) describe the dependence of galaxy properties in the different environments (Sect.~\ref{sec:prop_galaxies}).

\begin{figure}
    \centering
    \includegraphics[scale=0.45]{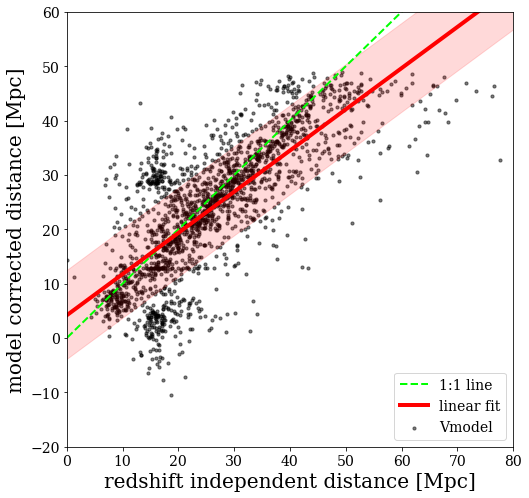}
     \caption{Comparison between the redshift-independent distances obtained from \cite{Steer2017} and those inferred from $v_{model}$ for the galaxies present in the \cite{Steer2017}'s catalog (black points). The green line is the 1:1 relation. The red line shows the linear fit to the point, while the shaded red region denotes the corresponding $\pm1\sigma$ scatter.}
    \label{fig:cfr_dist}
\end{figure}

\begin{figure*}
    \centering
    \includegraphics[scale = 0.55]{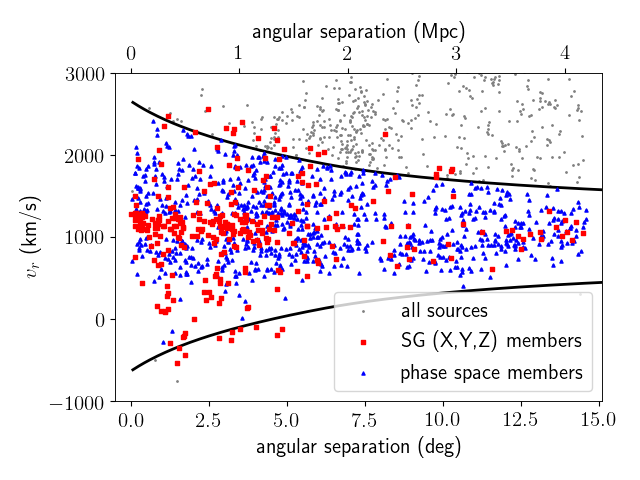}
    \includegraphics[scale = 0.45]{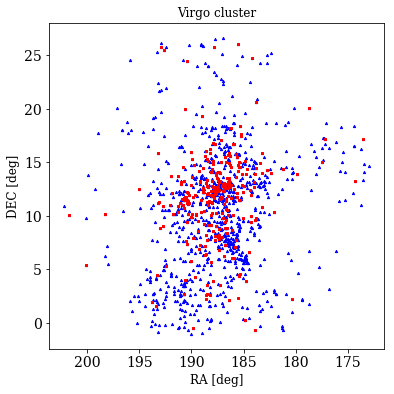}
    \caption{Left: phase space diagram for sources in the field of the Virgo cluster. The solid lines show the radial dependence of the escape velocity in the phase space diagram, according to the prescription by \citet{Jaffe2015}. Cluster members defined in the Super Galactic (X,Y,Z) coordinate frame (red points) or within the region (blue points) delimited by the caustics (i.e., the two solid lines) are distinguished from the remaining sources in the field of Virgo cluster (gray points). Right: projected distribution of the cluster members on the sky.  \label{fig:phase_space}}
\end{figure*}

\subsection{Cosmic distances}\label{sec:dist}
To characterize the positions of all galaxies around Virgo, we convert heliocentric velocities $v_r$ to intrinsic distances of the sources, according to the following steps.

First, we match our sample using the NED name to the NED-D catalog \citep{Steer2017}. 
The search yields a match for 1959 sources - corresponding to 29\% of the total sample - and for these sources in what follows we will adopt the \citet{Steer2017} distances ($D_{z-{\rm independent}}$) as the final cosmic distances. We calibrate all $D_{z-{\rm independent}}$ assuming $H_0 =$~74~km~s$^{-1}$~Mpc$^{-1}$ used in this work.  In the cases where the \citet{Steer2017} catalog provides multiple estimates for a given source, we  adopt the median distance. 

Second, we compute intrinsic distances following \citet{Mould2000}, using their method for correcting observed recession velocities for peculiar motions associated with various attractors in the local universe. We derive the correction $v_{\rm LG}$ of the observed heliocentric velocity of our galaxies to the centroid of the Local Group (LG), as in Eq.~A1 from \citet{Mould2000}. Then we estimate the correction $v_{\rm in,Virgo}$ that takes into account the infall towards Virgo attractor as in Eq.~1 by \citet[][]{Mould2000}. Distances and radial velocities relative to the Virgo center are calculated by means of the cosine theorem \citep[e.g.,][]{Karachentsev2010}. 
A cosmic velocity of $\sim1016$~km/s is assumed for Virgo,  as found in NED. It is obtained by correcting Virgo heliocentric velocity to the LG centroid for our infall velocity and for the infall of Virgo into the Great Attractor, as described in Appendix~A of \citet[][]{Mould2000}. 
We also assume a Virgo density profile $\rho(r)\propto r^{-2}$ and an amplitude $v_{\rm fid}=200$~km/s for the Virgo infall velocity \citep{Mould2000}. Model corrected velocities $v_{\rm model}$ are then derived as follows:
\begin{equation}
    \label{eq:vcosmic}
    v_{\rm model} = v_r+v_{\rm LG}+v_{\rm in,Virgo}
\end{equation}
Here we neglect higher order corrections in Eq.~A2 by \citet{Mould2000} that are due to the infall of our galaxies towards the Great Attractor and Shapley supercluster. We also assume a linear dependence between velocities and distances. 

Figure~\ref{fig:cfr_dist} shows the comparison between the model corrected distances ($D_{\rm model}$) and $D_{z-{\rm independent}}$ for 
galaxies with both redshift independent and model corrected distances. 
The median logarithmic difference is  $\log(D_{\rm model}/D_{z-{\rm independent}})=-0.03^{+0.12}_{-0.18}$. Here the reported uncertainties correspond to the 1$\sigma$ confidence interval. 
The comparison yields a negligible bias and an RMS scatter of $\sim0.1$~dex, which is consistent with that found in recent studies of the local Universe \citep{Leroy2019}. The small differences, well within the uncertainties, between these values and those reported in \citetalias{Castignani2021} are due to the different southern limit adopted in the two works, as here we do not consider galaxies at negative declinations as \citetalias{Castignani2021} did. 
Although we find an overall agreement between redshift independent and model corrected distances, we do see an increased dispersion in the data points in Fig.~\ref{fig:cfr_dist} along the y-axis at $\sim17$~Mpc, which corresponds to the distance of Virgo \citep{Mei2007}.  The model correction for Virgo galaxies is more uncertain, since peculiar velocities become significant as we approach the Virgo cluster.

Overall, from the comparison presented in Fig.~\ref{fig:cfr_dist} we conclude that for most of the galaxies in our sample with no redshift independent distance, the model-corrected version is reliable enough to determine galaxy 3D positions and local density estimates (see Sec.\ref{sec:ld}). The model corrected distances might not be as reliable for the Virgo cluster members, and in principle this could impact our estimates of local density. Nonetheless, we will show in Sect.~\ref{sec:prop_galaxies} that these uncertainties will not significantly affect our results.

To summarize, our adopted cosmic distances and velocities, $D_{cosmic}$ and $v_{cosmic}$, are the redshift independent distances and velocities, when available, and those derived as in Eq.~\ref{eq:vcosmic} for the remaining sources. 

We then make use of the Super Galactic (SG) coordinate system, which was developed by Gérard de Vaucouleurs. This coordinate frame has the equator aligned with the SG plane, which consists of a planar distribution of nearby galaxy clusters. The SG system is thus ideal for studies of the Cosmic Web in the local Universe. Therefore, assuming a linear relationship between $v_{cosmic}$ and distance, galaxies have been mapped into the Cartesian SG frame. In this frame galaxy positions are defined in terms of their SG coordinates SGX, SGY, and SGZ \citep{Tully2008}. We note that Virgo cluster center has (SGX; SGY; SGZ)~=~(-2.26; 9.90; -0.42)~$h^{-1}$~Mpc in the SG coordinate frame. 
At the coordinates of Virgo the SGY direction approximately corresponds to the line of sight.

\subsection{Membership to the Virgo cluster} \label{sec:cl}

\begin{table*}
\begin{adjustwidth}{-0.5cm}{}
\setlength\tabcolsep{2pt}
\caption{Detected filaments and spatial extent of their spines.\label{tab:filaments}}
\centering
{\small \begin{tabular}{|l|ccc|cc|c|}
\toprule
\multirow{2}{*}{Structure}  &   SGX & SGY & SGZ & RA & Dec.  & $L$ \\ 
   & [$h^{-1}$ Mpc] & [$h^{-1}$ Mpc] & [$h^{-1}$ Mpc] & (deg) & (deg) &  [$h^{-1}$ Mpc]\\ 
   \hline
   (1) & (2) & (3) & (4) & (5) & (6) & (7) \\ 
\hline
Leo Minor F. & $0.59\sim5.64$ & $4.06\sim6.30$ & $-2.83\sim-1.77$ & $120.27\sim160.23$ & $23.12\sim52.30$ & 7.63 \\
Canes Venatici F. & $0.64\sim3.97$ & $5.89\sim14.28$ & $1.35\sim4.79$ & $197.46\sim203.68$ & $34.47\sim44.13$ & 10.53\\
 Bootes F. & $5.9\sim10.59$ & $15.71\sim22.83$ & $5.91\sim11.51$ & $201.54\sim216.01$ & $43.31\sim60.89$ & 11.83\\
Ursa Major Cloud   &  $0.50\sim8.74$ & $2.67\sim13.95$ & $0.12\sim1.37$ & $177.62\sim186.07$ & $34.38\sim57.20$ & 15.44\\
LeoII B   F.   & $2.42\sim13.35$ & $13.65\sim13.97$ & $-8.37\sim-4.30$ & $131.00\sim163.99$ & $27.82\sim48.15$ & 12.67 \\
LeoII A   F.  & $0.43\sim9.18$ & $12.25\sim14.47$ & $-14.26\sim-6.73$ & $126.23\sim156.97$ & $15.79\sim33.13$ & 13.93 \\
VirgoIII   F.  & $-10.94\sim-5.51$ & $11.93\sim17.32$ & $3.40\sim11.09$ & $207.31\sim224.80$ & $2.32\sim5.42$ & 11.72 \\
Leo Minor B  F.  &$5.84\sim10.83$ & $18.28\sim21.53$ & $-7.01\sim-5.41$ & $152.99\sim163.15$ & $34.13\sim41.94$ & 7.95\\
W-M Sheet   & $-9.28\sim-3.41$ & $20.30\sim23.27$ & $-2.67\sim-1.91$ & $183.30\sim187.53$ & $1.59\sim15.15$ & 8.45\\
NGC5353/4  F. & $-12.57\sim9.42$ & $25.96\sim27.67$ & $0.30\sim9.20$ & $193.75\sim204.04$ & $2.09\sim47.84$ & 24.01\\
Serpens  F.   & $-5.48\sim-0.99$ & $11.02\sim17.01$ & $9.47\sim33.14$ & $230.39\sim256.6$ & $10.82\sim24.33$ & 25.63 \\
 Draco   F.  &  $13.98\sim18.77$ & $16.71\sim21.50$ & $14.90\sim22.51$ & $227.08\sim259.51$ & $58.76\sim60.92$ & 12.31\\
 Coma Berenices   F.  &$2.15\sim6.04$ & $12.89\sim38.31$ & $-4.92\sim-1.89$ & $173.06\sim175.40$ & $30.13\sim35.75$ & 26.27\\
\hline
\end{tabular}}
\end{adjustwidth}
Column description: (1) filament name; range in SG coordinates (2-4) and in projected space (5-6) spanned by the filament spine; (7) filament spine length \\
\end{table*}

To identify galaxies belonging to the Virgo cluster, we select galaxies within 3.6~$h^{-1}$~Mpc from the Virgo cluster center in the 3D Super Galactic coordinate frame. The chosen radius corresponds approximately to $\sim3r_{200}$, with $r_{200}=1.09$~$h^{-1}$~Mpc \citep{McLaughlin1999} the radius that encloses 200 times the critical matter density. 
The position of the { 311} Virgo members selected in this way in the phase space diagram is shown in Fig.~\ref{fig:phase_space}. 
Overall, they fall within the region  delimited by the caustics that are defined following the prescription by \citet{Jaffe2015}, assuming the $r_{200}$ radius and a concentration parameter of 2.8 as reported by \citet{McLaughlin1999}.
As the adopted definition is rather conservative, we also consider as cluster members those galaxies that fall within the cluster  region delimited in the phase space diagram by the caustics, regardless of their position in the Super Galactic coordinates. 
The final cluster member sample is the union of the members defined in Super Galactic coordinates and those defined using the phase space, for a total of { 1152} galaxies (526 above the magnitude completeness limit).

\begin{table*}
\begin{adjustwidth}{-0.5cm}{}
\setlength\tabcolsep{2pt}
\caption{Best fit parameters for the filament spines 
\label{tab:filament_fits}}
\centering
{\small \begin{tabular}{|l|c|c|c|c|}
\toprule
\multirow{3}{*}{Structure}  &   $\vec{a}$ & $\vec{b}$ & $\vec{c}$ & $\vec{d}$  \\ 
 & (a$_x$,a$_y$,a$_z$)  & (b$_x$,b$_y$,b$_z$)  & (c$_x$,c$_y$,c$_z$) & (d$_x$,d$_y$,d$_z$)  \\
  &   [$h^{-1}$~Mpc] & [$h^{-1}$~Mpc] &  [$h^{-1}$~Mpc] &  [$h^{-1}$~Mpc] \\ 
 \hline
 Leo Minor F. & -2.00 -2.00  2.00 & -5.00  -5.00      -4.32   &   11.59   6.52   1.38  &  0.59  4.54 -1.90 \\
 Canes Venatici F. &  2.00 -2.00  2.00 & -0.53 -5.00 -1.37   &   1.87  15.38    2.80  &  0.64  5.89  1.35 \\
 Bootes F. & -2.00 1.08 -2.00 & -5.00 -5.00 -5.00  & 2.62  11.03  12.28 & 10.28  15.71   5.91  \\
Ursa Major Cloud   & 2.00  -2.00  1.52 & 5.00         -5.00         -1.32  & 1.24  18.28   1.05 & 0.50  2.67  0.12 \\
 LeoII B   F.  & 2.00 -2.00  2.00 & 3.62  1.90  5.00  & 5.31   0.09 -10.28 & 2.42  13.66  -4.30  \\
 LeoII A   F. & -2.00 -2.00 -2.00  & -5.00 -5.00 -5.00  & 15.74   7.39  -0.54  &  0.43  12.25  -6.73 \\
VirgoIII   F.  & 2.00  2.00 -2.00 & -1.61  0.10 -5.00         &  -5.82   3.29  14.65 &  -5.51  11.93   3.40 \\
 Leo Minor B  F. & 2.00 -1.27  2.00  &  5.00   -5.00  -4.14 & -2.24  8.77  0.55 &  6.07  18.28    -5.42 \\
W-M Sheet  & 2.00  2.00 -2.00 &  5.00  5.00  1.52 & -1.20 -8.67  1.00 & -9.21  23.27  -2.67 \\
 NGC5353/4  F. & -2.00  2.00 -2.00 & -5.00          1.32 -4.48 & 28.99  -4.33  15.38 & -12.57  27.67    0.30 \\
 Serpens  F.   & 2.00 -2.00  2.00  & 5.00 -5.00  5.00 &  -2.89  12.74  16.66 & -5.10  11.02   9.47 \\
 Draco   F. & -2.00 -2.00 -2.00  & -5.00 -5.00 -5.00 & 11.25   2.50  14.56  & 13.98  21.21  14.90 \\
 Coma Berenices   F. & -2.00    2.00   1.59  & -5.00 3.46 -4.17   & 10.04 19.95 -0.45  & 2.15  12.89 -1.89 \\
\hline
\end{tabular}}
\end{adjustwidth}
Note. Each spine is parameterized with a polynomial curve $\gamma$:[0,1]~$\rightarrow\Omega$, such that $\gamma(t)=\vec{a}\,t^3+\vec{b}\,t^2+\vec{c}\,t+\vec{d}$. The best fit values of the parameters $\vec{a}$, $\vec{b}$, $\vec{c}$, and $\vec{d}$  are provided in (SGX, SGY, SGZ) coordinates for all filamentary structures considered in this work.
\end{table*}

\begin{table*}
\caption{A sample of filament points table. The full table is available online. \label{tab:general_properties_all_galaxies}}
\centering
\begin{tabular}{|c|c|c|c|c|c|c|c|}
\toprule
Filament & ID$_{point}$   & RA & DEC & SGX & SGY & SGZ & PA \\ 
  & & $[$deg$]$ & $[$deg$]$ & [$h^{-1}$ Mpc] & [$h^{-1}$ Mpc] & [$h^{-1}$ Mpc] & [deg] \\ 
\hline
  VirgoIII & 1 & 207.312 & 5.419 & -5.514 & 11.929 & 3.397 & 92\\
  VirgoIII & 2 & 207.875 & 5.398 & -5.572 & 11.961 & 3.543 & 92\\
  VirgoIII & 3 & 208.426 & 5.374 & -5.631 & 11.994 & 3.688 & 93\\
  VirgoIII & 4 & 208.966 & 5.348 & -5.69 & 12.027 & 3.831 & 93\\
  VirgoIII & 5 & 209.494 & 5.32 & -5.749 & 12.061 & 3.974 & 93\\
  VirgoIII & 6 & 210.012 & 5.289 & -5.809 & 12.094 & 4.116 & 94\\
  VirgoIII & 7 & 210.518 & 5.257 & -5.868 & 12.127 & 4.257 & 94\\
  VirgoIII & 8 & 211.014 & 5.223 & -5.928 & 12.16 & 4.397 & 94\\
  VirgoIII & 9 & 211.499 & 5.187 & -5.989 & 12.194 & 4.535 & 94\\
  VirgoIII & 10 & 211.973 & 5.15 & -6.049 & 12.227 & 4.673 & 95\\
 ... & ... & ... & ... & ... & ... & ... & ... \\ 
   Draco & 1 & 227.076 & 58.758 & 13.983 & 21.212 & 14.895 & 67\\
   Draco & 2 & 227.476 & 58.843 & 14.095 & 21.237 & 15.04 & 68\\
   \hline
\end{tabular}
\begin{flushleft}
Column description: (1) filament name; (2) integer index associated with the curve parameter $t$; (3-4) projected coordinates, (5-7) SG coordinates, and (8) position angle for the filament spine at the index reported in column (2).
\end{flushleft}
\end{table*} 

\subsection{Filamentary structures} \label{sec:fil}

Moving beyond the cluster, we aim to characterize its surrounding Cosmic Web in 3D.
We start by considering the eight filamentary structures presented in \cite{Tully1982,Kim2016}: the W-M Sheet located to the south of Virgo; the nearby Ursa Major cloud in the North; the  VirgoIII filament to the south of Virgo; the extended NGC5353/4 filament, with the corresponding group at the end of it; the Canes Venatici filament just north of  NGC5353/4 filament; and the LeoII~A, LeoII~B, and Leo Minor filaments belonging to the Leo cloud to the northwest. To test the reliability of these filaments, we construct a series of different SGX–SGY–SGZ volume slices with an arbitrary depth of 4~$h^{-1}$~Mpc along the SGY axis. Selected structures are confirmed by visual inspection of the SGX–SGZ projection of each slice, looking for overdense and long (i.e., filamentary) galaxy distributions. All candidate structures are  present in consecutive slices. During this visual inspection of the distribution of galaxies in the SGX-SGZ plane, we identify five additional structures that  were not reported in \citet{Kim2016} and that will enter our final filament sample. We name these respective structures the Leo Minor B, Bootes, Serpens, Draco, and Coma Berenices filaments, where the names of these filaments derive from the dominant constellation that they are in.  As further discussed in Sect.~\ref{sec:comparison_filament_works}, all these structures have a least one main counterpart in the V8k catalog of nearby sources and structures \citep{Courtois2013}.

Nevertheless, we stress that the goal of this work is not to provide a complete census of all filaments around Virgo.  Instead, we provide a detailed characterization of the filaments that have the highest density contrast relative to the surrounding field as determined by visual inspection, including those already known in the Northern hemisphere.

The 13 structures all fall within the cuboid enclosed by the following limits:
\begin{eqnarray}
\rm -13 < SGX / (h^{-1}~Mpc) < 20, \nonumber \\
\rm 2 < SGY / (h^{-1}~Mpc) < 38, \nonumber \\ 
\rm -15 < SGZ  / (h^{-1}~Mpc) < 33. \nonumber
\end{eqnarray}
These limits correspond to a more extended region in the Northern Hemisphere than that considered by \cite{Kim2016}, which allows us to have a more comprehensive characterization of the large-scale structures around Virgo than previous  studies. 
Note that the (SGX; SGZ) coordinate frame approximately corresponds to the plane of the sky where filamentary structures are better defined, while the SGY axis is associated with the line of sight, and thus more impacted by positional errors arising from distance uncertainties.

Similar to \citet{Kim2016}, for each filamentary structure we consider an associated parallelepiped $\Omega$ in the 3D SG frame, large enough to conservatively enclose all galaxies that belong to the structure. We set the parallelepiped dimensions after the visual inspection of the filamentary structure in (SGX,SGY,SGZ) coordinates, with different projections.
We then determine the filament spines by fitting the locations of the galaxies in SG coordinates. 
We parameterize the spine of each filament by fitting a third-order polynomial curve   $\gamma$:[0,1]~$\rightarrow\Omega$, such that $\vec{\gamma}(t)=\vec{a}\,t^3+\vec{b}\,t^2+\vec{c}\,t+\vec{d}$. Here $\vec{a}$, $\vec{b}$, $\vec{c}$, and $\vec{d}\in\mathbb{R}^3$ are the curve parameters with their origin coincident with the Sun, as this is the case for the SG (X,Y,Z) coordinate system.We then perform a fit by minimizing the sum of the distance squares of each galaxy in $\Omega$ to the filament spine. 
In Table~\ref{tab:filaments} we report the spatial extent of the filaments, which span a wide range in length, between $L\sim(8-26)~h^{-1}$~Mpc.  
In Table~\ref{tab:filament_fits} we report the best fit parameters of our fits to the filament spines. In Table~\ref{tab:general_properties_all_galaxies} we provide different points that sample the filament spines both in projection and in the 3D SG frame. In the same Table we also report the position angles of the tangent vectors, along each filament spine.

\begin{table*}
\setlength\tabcolsep{2pt}
\caption{Number of galaxies in each filament. Column description: (1) structure; (2) total number of galaxies, regardless of their magnitude (N$_{gal}$); (3) number of galaxies above the survey magnitude limit (N$_{gal}$@$M_{r_{lim}}$); (4) magnitude limit of each structure; (5) number of galaxies above the magnitude limit given in (4); (6) Number of galaxies in the \citet{Kim2016} sample. \label{tab:filaments_mag}}
\centering
 \begin{tabular}{|l|c|c|c|c|c|}
 \toprule
Structure  & N$_{gal}$ &  N$_{gal}$@$M_{r_{lim}}$ & $M_{r_{f, lim}}$ & N$_{gal}$@$M_{r_{f, lim}}$ & K16  \\ 
\hline
Virgo Cluster & { 1152} & 526 & -15.41 & { 570} & \\
 Leo Minor  F. & { 124} & 13 & -12.86 & { 62}  &54 \\
 Canes Venatici  F. & 96 & { 24} & -14.01 & 48 & 51 \\
 Bootes F.  & 169 & { 113} & -15.07 & { 136} &  \\
Ursa Major Cloud & { 580} & 117 & -14.00 & 217  & \\
 LeoII B  F.& 63 & 28 & -14.52 & 43 & 105 \\
 LeoII A  F.& 145 & 53 & -14.64 & 97 &180\\
VirgoIII  F. & 206 & 115 & -14.69 & 148 & 181 \\
 Leo Minor B  F.& 39 & 28 & -14.90 & 29 &\\
W-M Sheet  & { 345} & { 198} & -14.96 & 250 & 256 \\
 NGC5353/4  F.& 133 & 90 & -15.34 & 106 & 102 \\
 Serpens  F.& 65 & 34 & -15.35 & 39 &\\
 Draco  F.& 48 & 44 & -15.61 & 45 & \\
 Coma Berenices  F. & 105 & { 62} & -15.32 & 69 & \\
pure field & { 2249} & { 1160} &-&- &\\
poor groups & { 1086}  & { 652} &-&- &\\
rich groups & { 1626} & { 937} &-&- &\\
all & { 6780} & { 3528}  & - & - &\\
\hline
\end{tabular}

\end{table*}

To identify filament members we select galaxies found within $2~h^{-1}$~Mpc of the spine, with the radial cut selected to minimize the contamination from the field \citep{Lee2021,Galarraga2020}. We verified {\it a posteriori} that all considered filamentary structures are overdense and elongated over several Mpc in length. Indeed, as further outlined in Sect.~\ref{sec:density_profiles}, the density contrast, evaluated as the ratio between the average number density of galaxies within $1~h^{-1}$~Mpc from the filament spine relative to the field value, ranges between $\sim3-18$, which thus strengthens the reliability of the selected filaments.

\begin{figure}[h!]
    \centering
   \includegraphics[width=.5\textwidth]{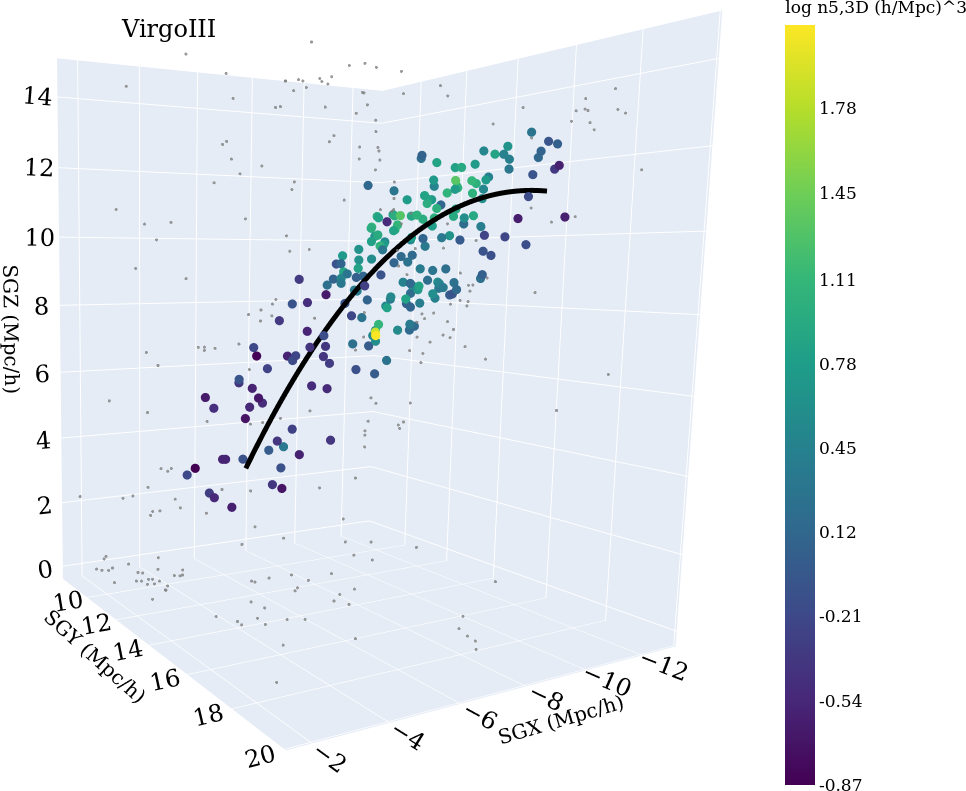}
    \caption{Galaxies in the vicinity of the Virgo~III filament. Galaxies within 2~Mpc are color-coded by the 3D local density, and galaxies with separations greater than 2~Mpc are shown with the grey points. The filament spine is shown with the black curve.  Interactive 3D plots for all filaments are available in the electronic edition of this article.}
    \label{fig:virgoIII-3D}
\end{figure}

\begin{figure*}
    \centering
    \includegraphics[scale = 0.6]{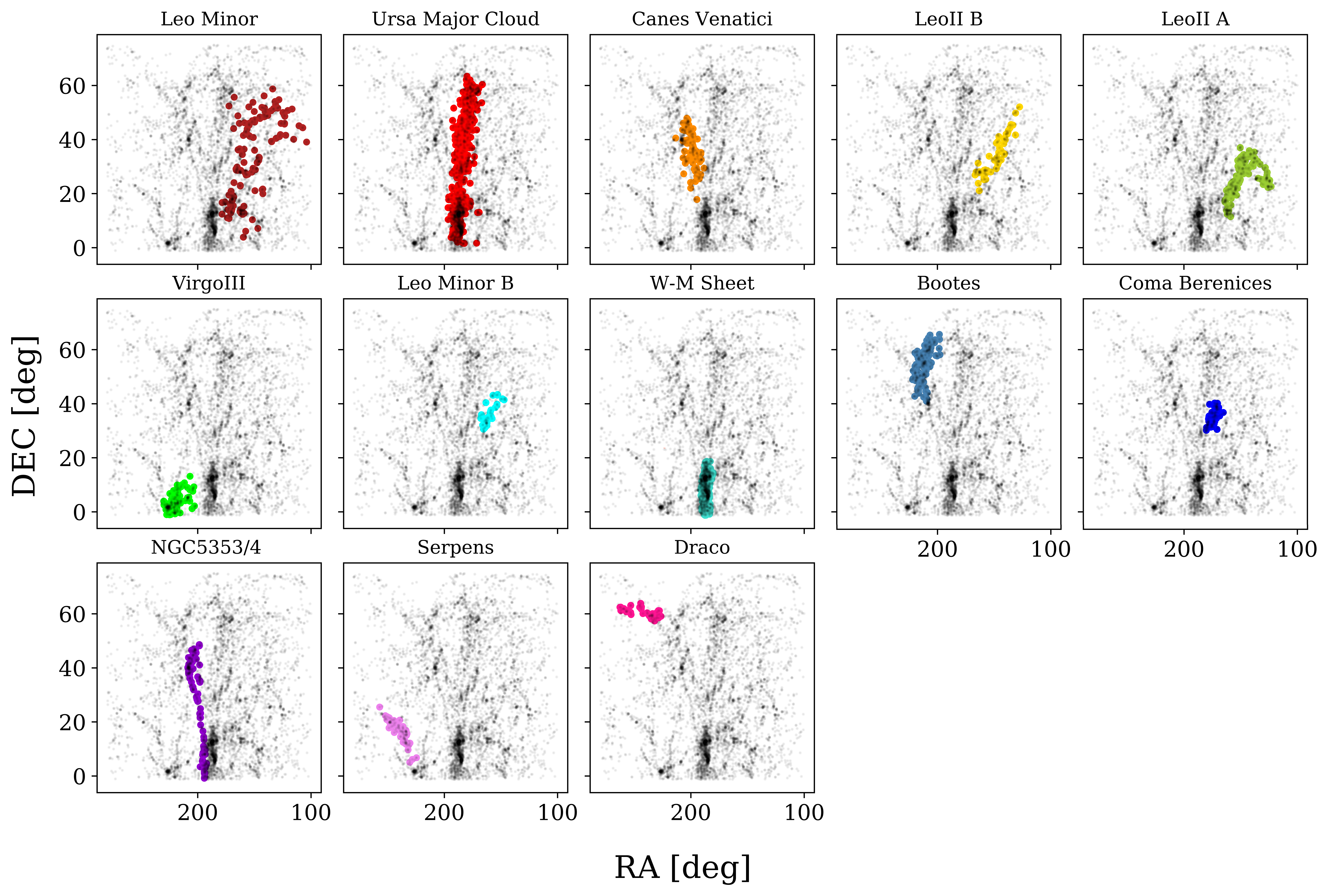}
    \caption{Spatial distribution of galaxies around the Virgo galaxy cluster. Gray points show all  galaxies, colored points show galaxies belonging to the different filaments. This color scheme will be kept in all plots.
    \label{fig:positions_fil}}
\end{figure*}

As an example of the outcome of our procedure, Figure \ref{fig:virgoIII-3D} shows the selected parallelepiped in the SG frame within which the VirgoIII filament is embedded. The filament spine and filament members within 2~$h^{-1}$~Mpc from the spine are highlighted. The latter are color-coded by their local density (see Sect.\ref{sec:ld}) to highlight density variations along the filament. These variations are also due to the presence of groups within the filament (see Sect.\ref{sec:other_envs_gr}). The complexity of these structures motivates further characterization of environment, even within filaments.  

Figure \ref{fig:positions_fil} shows the spatial distribution of the { 2118} galaxies belonging to the identified structures. Filaments are sorted by increasing distance from us, and this color scheme is adopted throughout the paper to help the reader track the different filaments. 
It appears evident that different filaments exhibit different properties in terms of their distance, richness, and structure. In particular, the W-M sheet has a planar morphology, as further discussed in the following Sections. The Leo filaments were originally classified as a single cloud by \citet{Tully1982}. Furthermore, the Ursa Major Cloud and the W-M sheet overlap with the  Virgo cluster periphery.
Indeed, { 418} cluster galaxies are also members of the Ursa Major Cloud { (214)} or the W-M Sheet { (204)}. 
This is primarily due to the difficulty in unambiguously distinguish Virgo cluster members from those of nearby correlated structures, as further discussed in previous studies \citep[e.g.,][]{Kim2014,Kourkchi2017}.

\begin{figure}
    \centering
    \includegraphics[scale = 0.3]{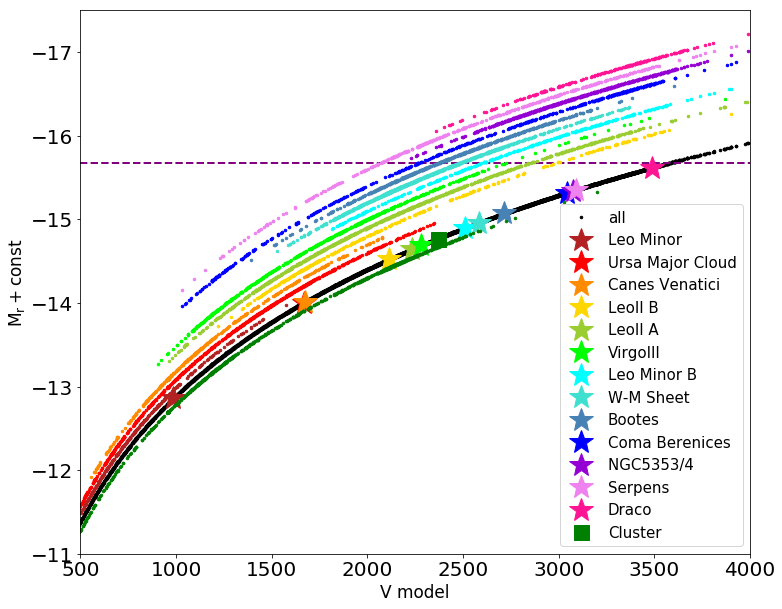}
    \caption{Absolute magnitude  $\rm M_r$ as a function of velocity  V$_{\rm model}$ (black line) and magnitude limit of the survey (horizontal dashed line). Stars represent the magnitude limit proper of each filament separately, the green square the magnitude limit of the cluster. These limits are obtained as the the magnitude including 95\% of the data. For display purposes, points of the different filaments are also shown in colors, with an arbitrary vertical shift to avoid the superimposition of the points. 
    \label{fig:mag_lim}}
\end{figure}

Each filament is located at a different mean distance, and the very conservative absolute magnitude limit for the catalog was set by the most distant galaxies in the entire sample.  We therefore compute an absolute magnitude limit that is appropriate for each filament (Fig.\ref{fig:mag_lim}), that might be useful for specific studies (e.g. comparing the local density of filament galaxies with the density in the surrounding field, see Fig.~\ref{fig:distribution_density_fil}). Specifically, we compute the completeness limit by computing the distance modulus from the distance encompassing 95\% of galaxies in any given filament.
Table \ref{tab:filaments_mag} reports the magnitude limit and the number of galaxies above it for each filament. 

\subsubsection{Comparing different filament determinations}\label{sec:comparison_filament_works}

As we follow the approach presented by \cite{Kim2016}, we now briefly compare our results with theirs. 
In Table~\ref{tab:filaments_mag} we report the number of filament members identified by \cite{Kim2016}, for the seven filaments in common. For these filaments the ratio of the total number of members found in this work to that reported by \citet{Kim2016} ranges between $\sim0.6-2.3$, with a median value of 1.3. This wide range of values is due to both the different input catalogs used (our catalog has been carefully cleaned of duplicates and includes sources from additional surveys) and to the different filament membership assignments.

Memberships to filaments around the Virgo cluster can be retrieved also from the \cite{Tempel2014_f} catalog. However, our approach have been fine-tuned to characterize filaments specifically around Virgo,  whereas \cite{Tempel2014_f} have searched for a large sample of filaments in the Sloan Digital Sky Survey (SDSS) over a wider field and up to larger distances (up to 450~$h^{-1}$~Mpc). 
Their method approximates the filamentary network using a random configuration of small segments (thin  cylinders). 
If we cut the \citet{Tempel2014_f} catalog at our velocity limit ($z<0.012$), we retain only 1281 galaxies, and 774 of these are associated to 39 filamentary structures made of more than 10 galaxies.  This includes filaments in the location of the Serpens, Bootes, Canes Venatici, NGC5353/4 filaments, the Ursa Major Cloud, and the W-M Sheet, but these filaments have many fewer members. Overall their filaments are much less populated: the median number of filament members found within 1~$h^{-1}$~Mpc from the filament spine in 15. It appears therefore that to carefully characterize filaments in the local Universe, it is not appropriate to apply a general approach that is optimized for a much larger ($z<0.155$) redshift range. The above considerations motivated us to exploit a method that is tailored to the case of the local Universe and specifically to filaments around Virgo. 

 In this context, it is worth mentioning the V8k catalog of nearby sources and structures that is discussed by \citet{Courtois2013} and is part of the Extragalactic Distance Database \citep{Tully2009}\footnote{http://edd.ifa.hawaii.edu/index.html}. This catalog provides a census of large scale structures in the local Universe and their members. By cross matching, via membership, the structures considered in this work with those listed in the V8k catalog we found that all of our structures have at least one main counterpart in V8k. This also applies to the 5 filamentary structures mentioned above, that are not in \citet{Kim2016}, but are considered in this study. Indeed, Leo Minor B of this work is mostly matched with the Crater Cloud in V8k; Serpens with the Serpens Cloud; Draco with the Bootes cloud; Coma Berenices with the Ursa Major Southern Spur. Bootes filament has two main counterparts in V8k: the Bootes Cloud and the Canes Venatici - Camelopardalis Cloud. 

By matching the V8k galaxy catalog with ours we also found 232 galaxies that  belong to V8k structures that are not matched to any of ours, namely Cancer - Leo Cloud, Draco Cloud, V8k structure ID 322, and Ophiuchus Cloud.  These structures are located along the periphery of the field around Virgo considered in our study.  The presence of these clouds is not a major concern for our study. They only marginally contaminate our field sample, as in fact 128 out of the 232 (i.e., 5.6\%) are classified as pure field galaxies in our work (see Sect.~\ref{sec:other_envs_gr}). 

\subsubsection{Radial Density Profiles}\label{sec:density_profiles}
\begin{figure*}
    \centering
    \includegraphics[trim = 0 130 0 0, width=0.9\textwidth]{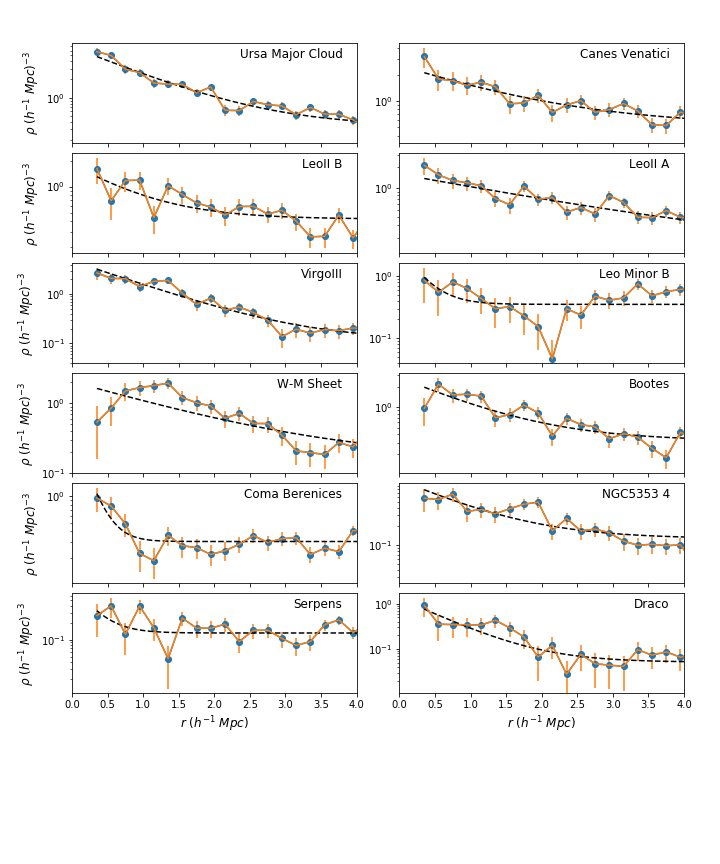}
    \caption{Galaxy density versus distance from the filament spine.
    Dashed lines show the exponential fit, see text for further details. }
    \label{fig:density-profiles}
\end{figure*}

In this section we provide an estimate of the 
width and density contrast of the filaments, with the goal of better characterizing these overdense structures. Following \citet{Lee2021}, we investigate how the number density of filament galaxies depends on the distance to the spine, and we calculate the density of galaxies in cylindrical shells as a function of 3D distance from the filament spine. Average densities $\rho$  at a distance $r$ from the filament spine are calculated within cylindrical volumes $V = \pi L[ (r+\delta r)^2 -(r-\delta r)^2]=4\pi L\,\delta r$ as
\begin{equation}
\label{eq:rho_r}
\rho(r) = \frac{N_{gal}(<r+\delta r)-N_{gal}(<r-\delta r)}{4\pi L\delta r}\,,
\end{equation}
where $L$ is the length of the filament (see Table \ref{tab:filaments}).  We increase the radius  from 0.2 to 6~$h^{-1}$~Mpc in increments of 0.2~$h^{-1}$~Mpc, while we choose $\delta r = 0.1$~$h^{-1}$~Mpc.
We show the resulting density profiles in Fig.~\ref{fig:density-profiles}.  
The filaments span a range of densities (y-range of individual plots varies to improve readability).  When comparing densities within  $\sim1~h^{-1}$~Mpc from the spine, the Ursa Major Cloud is the densest filament, and the Serpens Filament is the least dense.  

Almost all profiles show a decrease in galaxy density as distance increases.  The profiles describing some filaments flatten out at $r>3~h^{-1}$~Mpc (e.g. the Leo B, Coma Berenices, Leo Minor B), while others continue to decline over the full range of the radii probed (e.g. the Ursa Major Cloud, and VirgoIII). 
These results suggest that the region around the filament spine is indeed where the clustering of galaxies is stronger, and they strengthen the characterization of the filament skeletons adopted in this work. 

The W-M sheet is an exception: it appears to be the only structure for which the density is not clearly declining with distance.
Omitting the first two points at small radii that have large uncertainties due to small number statistics, the profile is fairly flat up to 1.5~$h^{-1}$~Mpc,  and declines at larger distances. This finding is not surprising, given the planar distribution of galaxies in this structure \citep[e.g.,][]{Kim2016}.

\begin{table}
\begin{adjustwidth}{-0.5cm}{}
\setlength\tabcolsep{2pt}
\caption{Best fits for the density profiles of the filaments, based on the whole sample.
\label{tab:density_profile_par}}
\centering
{\small \begin{tabular}{|l|c|c|c|c|}
\toprule
\multirow{2}{*}{Structure}  &   a &  b & $r_0$  \\ 
        &   [$h^{3}$~Mpc$^{-3}$] & [$h^{3}$~Mpc$^{-3}$] & [$h^{-1}$~Mpc] \\ 
 \hline
 Leo Minor F. &  $2.02\pm0.32$ & $0.15\pm0.26$ & $2.67\pm1.24$ \\
 Canes Venatici F. & $2.05\pm0.43$ & $0.52\pm0.05$ & $1.40\pm0.31$  \\
 Bootes F. & $2.44\pm0.74$  & $0.32\pm0.03$ & $0.91\pm0.21$ \\
Ursa Major Cloud   & $6.54\pm0.99$ & $0.34\pm0.03$ & $0.90\pm0.09$  \\
 LeoII B   F.  & $1.39\pm0.90$ & $0.42\pm0.03$ & $0.76\pm0.38$  \\
Virgo LeoII A   F. & $1.48\pm0.21$ & $0.12\pm0.08$ & $2.21\pm0.59$ \\
VirgoIII   F.  & $4.66\pm0.71$ & $0.11\pm0.02$ & $0.87\pm0.08$  \\
 Leo Minor B  F. &  $2.37\pm8.89$ & $0.35\pm0.04$ & $0.26\pm0.51$ \\
W-M Sheet & $1.89\pm0.49$ & $0.16\pm0.06$ & $1.42\pm0.36$  \\
 NGC5353/4  F. & $1.00\pm0.47$ & $0.13\pm0.02$ & $0.83\pm0.30$  \\
 Serpens  F. & $0.79\pm1.84$ & $0.13\pm0.01$ & $0.25\pm0.30$  \\
 Draco   F. & $1.27\pm0.64$ & $0.05\pm0.01$ & $0.60\pm0.17$  \\
 Coma Berenices   F. & $5.44\pm7.56$ & $0.31\pm0.01$ & $0.18\pm0.10$  \\
\hline
\end{tabular}}
\end{adjustwidth}
Note. The density profile is parameterized as $\rho(r) = a\exp\left(-\frac{r}{r_0}\right)+b$.
\end{table}

\begin{figure*}
    \centering
    \includegraphics[width=.45\textwidth]{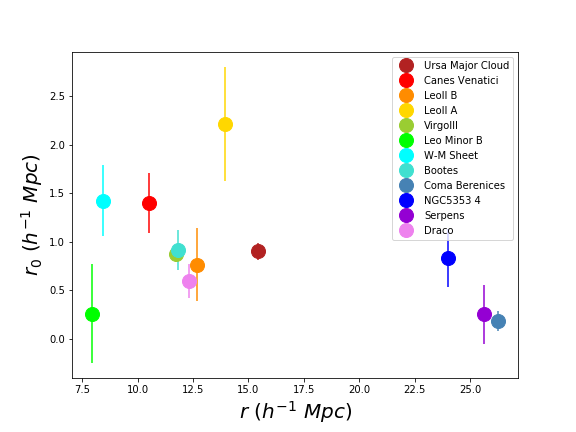}
    \includegraphics[width=.45\textwidth]{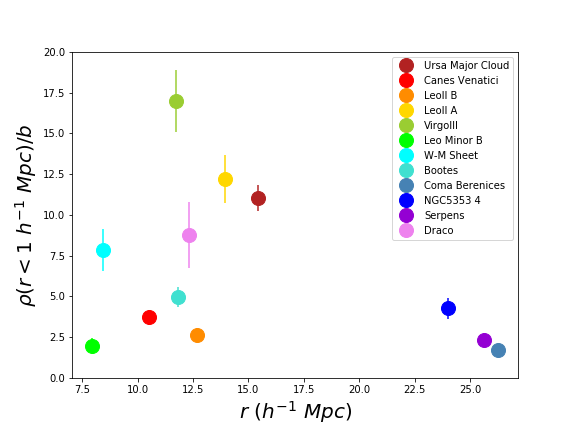}
    \caption{Scale length $r_0$ (left) and central density contrast (right), i.e., the ratio of the density enclosed within $1~h^{-1}~{\rm Mpc}$ to the best fit  value $b$ at large radii, as a function of filament length.}
    \label{fig:scale_length_contrast}
\end{figure*}

We fit the density profiles of each filament as a function of perpendicular distance from the spine, $r$, with an exponential law:
\begin{equation}
\rho(r) = a\exp\left(-\frac{r}{r_0}\right)+b\,,
\end{equation}
where: 
$a$ is the best-fit central density at $r\ll r_0$, above the field value; $b$ is the best fit for the field density at large scales $r\gg r_0$; and $r_0$ is the exponential scale width of the filament.  The best-fit parameters are reported in Table~\ref{tab:density_profile_par}, while Fig.~\ref{fig:scale_length_contrast} shows the exponential scale width $r_0$ and the central density contrast as a function of filament length. The central density contrast is defined as the density enclosed within $r<1~h^{-1}$~Mpc divided by the best fit value of $b$. On average we find  $r_0=(0.9\pm0.7)~h^{-1}$~Mpc. We report here the median value along with the rms dispersion around the median.\footnote{Note that $r_0$ is overall smaller than the value of $2.3~h^{-1}$~Mpc that we found in \citetalias{Castignani2021}. Discrepancies might be due to both the different sample selection and the different local density estimator adopted. 
Indeed, \citetalias{Castignani2021} considered only filament galaxies in a mass complete sample and the 5th nearest-neighbor density estimator. This is a good proxy for the local density, but overdense and underdense substructures within the filaments tend to increase the scatter of $n_5$ when plotted vs $r$. On the other hand, the density in Eq.~\ref{eq:rho_r} is averaged in cylindrical shells, so that this observed scatter is limited.}
Interestingly, the long filaments tend to have small values of $r_0<1$~$h^{-1}$~Mpc and low density contrasts $<5$, whereas shorter filaments with $L<17$~$h^{-1}$~Mpc have a larger dispersion and reach higher values for both $r_0$ and the density contrast.

This analysis is based on the full catalog of $\sim7000$ sources. This allows us to better recover the structural parameters of the filaments with maximum signal-to-noise ratio. If we repeat the analysis using the magnitude-limited sample, we obtain similar results but strong shot noise in several radial bins prevents us from deriving robust fits.
By using the full catalog we might be biased towards observing the highest number densities for the nearest filaments. For example the Ursa Major Cloud is nearby and very rich. Similarly, other closer filaments such as Leo Minor and Canes Venatici show  high central densities. However, our key estimated parameters such as the density contrast and the scale length $r_0$ are fairly independent from the exact galaxy selection, as they are determined relative to the field density value, which is set at large radii $r\gg r_0$.

Our results are consistent with the theoretical expectations of \citet{Galarraga2020} for the local Universe, who find that long filaments are thinner and less dense than shorter ones.  Compared to the best fits by \citet{Lee2021} for the major VirgoIII, Canes Venatici, LeoII A, LeoII B, Leo Minor, and NGC 5353/4 filaments, we find smaller central densities and higher scale length parameters. They found  $r_0<1~h^{-1}$~Mpc for all their filaments, which may be due to the fact that they adopted a different approach. In particular, they used a moving bin along the radial direction to estimate the density and fit the profile fixing $b=0~h^{3}$~Mpc$^{-3}$.

\section{Additional environmental metrics} \label{sec:other_envs}

\subsection{Groups and field around Virgo} \label{sec:other_envs_gr}

In Section \ref{sec:fil} we focused on the determination of the filaments, neglecting the presence of other structures, e.g. galaxy groups. It is likely that groups are present both within filaments and in other field regions. As a consequence, galaxies outside of the Virgo cluster or the identified filaments are not necessarily purely field galaxies. 
To identify galaxy groups within our sample, we match our catalog to the environmental catalog from \citet{Kourkchi2017}.  They characterized galaxy groups in our immediate neighborhood ($v_r<3500$ km/s). Their group finding procedure starts with the most luminous galaxy and iteratively associates galaxies that fall within its turnaround radius.  The algorithm then proceeds to the next most luminous galaxy that is not already assigned to a group, and the process repeats. Their galaxy catalog involves a compilation of sources taken from the Lyon-Meudon Extragalactic Database (LEDA\footnote{http://leda.univ-lyon1.fr/}), the 2MASS Redshift Survey, 2MRS11.75 \citep{Huchra2012}, and NED. 
For each galaxy in their catalog, \citet{Kourkchi2017} provide the membership to a group and the properties of the group. Of interest for our scope is the halo mass of the hosting structure, derived from the Ks-band luminosity by using M/L ratios given in their Eq. 8.
Therefore, \citeauthor{Kourkchi2017}'s catalog allows us (1) identify galaxies that, regardless of their membership to any filament,  belong to a group; (2) obtain a ``clean'' pure field sample made up of galaxies not belonging to any filaments nor associated with groups of two or more galaxies; and (3) obtain a halo mass estimate of the hosting structure for each galaxy in the sample. 

We cross-match our galaxy catalog and the catalog of group galaxies of \citeauthor{Kourkchi2017} using a search radius of 10\arcsec and we find { 5651} matches (83\% of the sample). For the { 1129} galaxies with no match in the Kourkchi catalog, we assign the group membership of their closest neighbour in 3D space.

We then classify as pure field galaxies those that are isolated based on \citeauthor{Kourkchi2017}'s classification and do not belong to the Virgo cluster or to any filament.  { 2249} galaxies in our catalog ({ 1160} above the magnitude completeness limit) are pure  field galaxies. Regardless of their membership in any filaments,  { 1086} ({ 652} above the magnitude limit) galaxies belong to groups with $2\leq N_{mem}<5$, with $N_{mem}$ being the number of group members identified in the \citeauthor{Kourkchi2017} catalog. Hereafter, we refer to the 2$\leq N_{mem}<5$ groups as {\it poor groups}.  We define rich groups as those with $N_{mem}\geq 5$, and we find that { 1626} galaxies (937 above the magnitude limit) belong to a rich group and are not in the Virgo cluster. The median (mean) number of members in a group is 8 (15).

\begin{figure}
    \centering
    \includegraphics[scale=0.35]{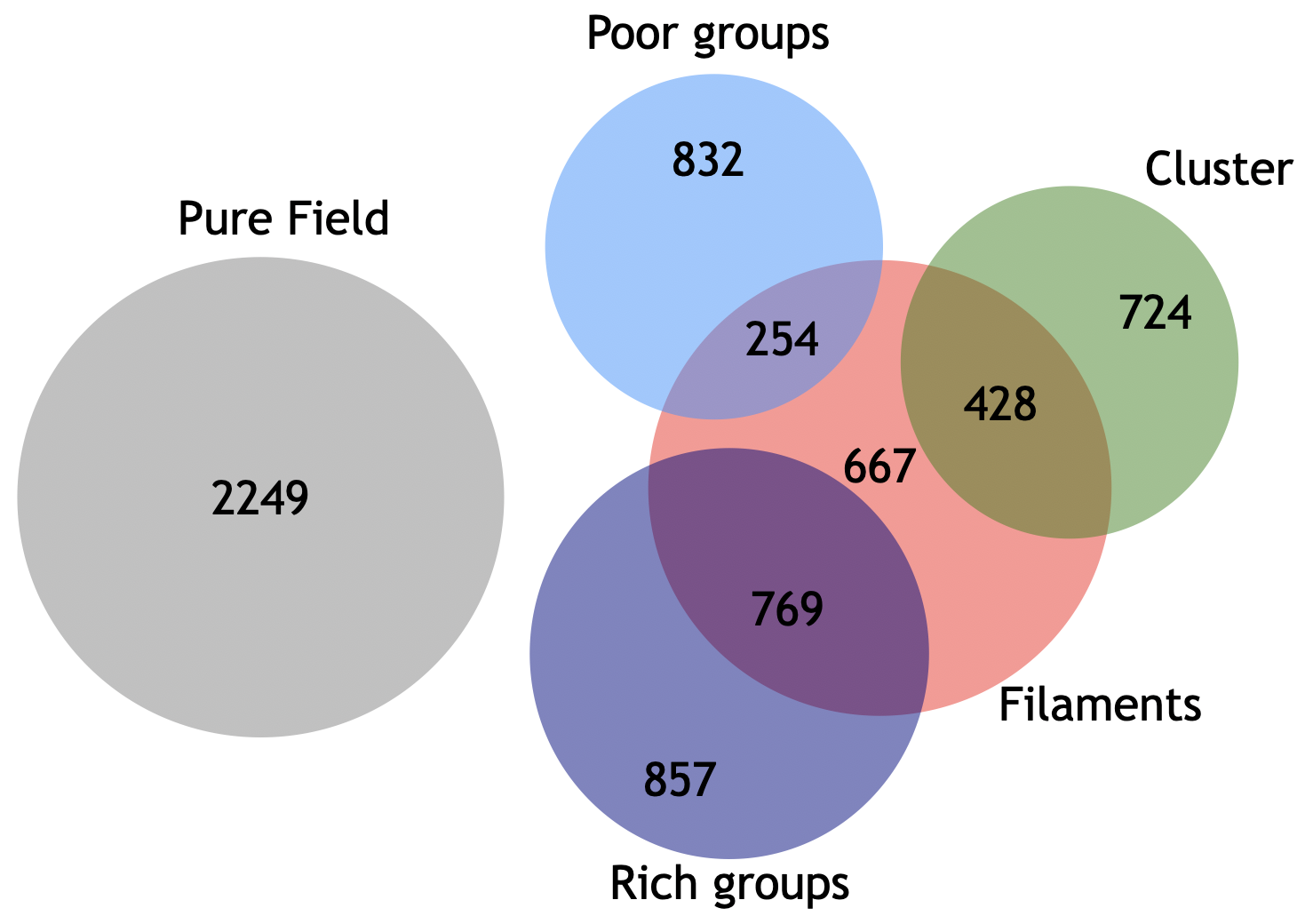}
    \caption{Euler-Venn diagram summarizing the distribution of galaxies in the different global environments.  \label{fig:venn_environment}}
\end{figure}

Figure \ref{fig:venn_environment} summarizes the different environments considered, showing the  overlap among the different classes. A significant fraction (33\%) of  galaxies in our sample are pure field galaxies, while the remaining ones are associated with Mpc-scale overdense structures: the Virgo cluster (17\%), the surrounding filaments (31\%), and groups (40\%). Filaments are a very heterogeneous environment: 20\% of their galaxies are in common with Virgo and are thus classified as members of both the cluster and a filament (Sect.~\ref{sec:fil}), 12\% of them are also located in poor groups, and 36\% of them are also found in rich groups.  It is therefore essential to distinguish among the different global environments in which sources live if we are to understand the impact of these environments on the observed properties of galaxy.

We then extract from \cite{Kourkchi2017}'s catalog the halo mass of the hosting system.
We note that $\sim 20\%$ of our cluster galaxies are not  members to Virgo according to \cite{Kourkchi2017} but instead are formally associated with lower mass halos, with masses uniformly distributed down to $\log(M_{\rm halo}/M_\odot)\sim10$. This discrepancy is due to differences in the cluster membership assignments between \citet{Kourkchi2017} and this work, in particular in the outskirts of the Virgo, where the memberships are more uncertain.  To avoid confusion and to be consistent with the Virgo membership definition used in this paper, we assign them the halo mass of Virgo $\sim10^{15}~M_\odot$ \citep{Fouque2001,Kourkchi2017}.

\subsection{Local density} \label{sec:ld}

\begin{figure}
    \centering
    \includegraphics[scale=0.6]{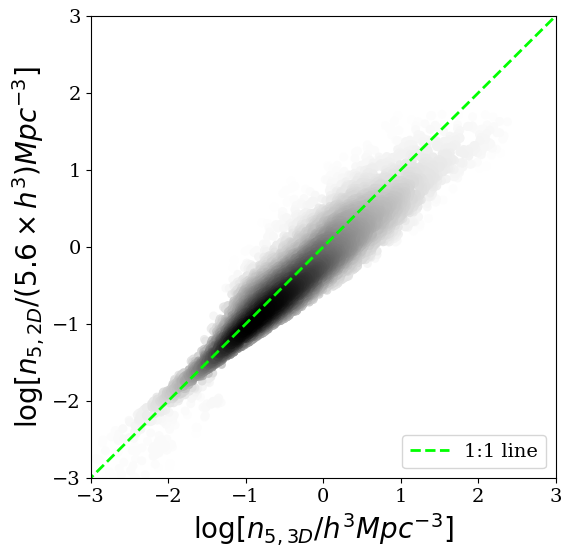}
    \caption{Comparison between the local number density  estimates for the galaxies in the sample. The x-axis shows the 3D volume number densities. The y-axis displays the 2D surface densities translated into a 3D volume densities, obtained dividing by the slice width $\Delta {\rm SGY}=5.6~h^{-1}$~Mpc. 
    \label{fig:cfr_density}}
\end{figure}

In the previous sections we focused on a global parameterization of the environment. We now focus on a more local prescription in terms of local density.
For each galaxy in the catalog, we compute $k$-nearest neighbor density (with $k = 5$)\footnote{Using other estimators, such as the modified 10-th nearest neighbor density \citep{Cowan2008}, will not affect the results.}. 
This is a widely used non-parametric estimate for the local environment of galaxies that is largely independent of the dark matter halo mass \citep[see e.g.,][for a review]{Muldrew2012}.
We consider only neighbors in the catalog whose r-band absolute magnitude is  $M_r \leq -15.7$, the completeness limit of the survey, to avoid biasing our estimates towards lower values at higher distances.\footnote{Note that in \citetalias{Castignani2021} we have not applied a magnitude cut to compute local densities. Therefore the two measurement of local density are not directly comparable.}

Specifically, local densities are computed in 3D ({\it volume densities}) in the  
(SGX,SGY,SGZ) Cartesian frame and in 2D  ({\it surface densities}) by projecting separations onto the (SGX, SGZ) plane.  The 2D density is evaluated by including galaxies within a $\Delta$SGY$=5.6~h^{-1}$~Mpc width, that corresponds to the $2\sigma$ statistical uncertainty along the line of sight at the distance of Virgo (see Fig.~\ref{fig:cfr_dist}). 
As outlined in Sect.~\ref{sec:dist} line of sight uncertainties are in fact of the order of $\sim0.1$~dex and may affect our 3D analysis. 

To investigate possible biases in the density estimates, we compare the 2D vs 3D local densities in Fig.~\ref{fig:cfr_density}. The two density estimates are consistent with each other once the 2D estimates are rescaled for the SGY width to convert them into 3D densities, i.e by dividing them by 5.6 $h^{-1}$~Mpc. The median logarithmic difference $\log(n_{5,3D}) - \log(n_{5,2D}/\Delta{\rm SGY}) = 0.06^{+0.29}_{-0.28}$ yields a negligible bias, well within the reported $1\sigma$ confidence interval. Given that results  obtained with the 2D and 3D local density estimates are quantitatively in agreement, from now on we will be considering only the 3D densities.

As previously shown in Fig.~\ref{fig:phase_space}, velocity dispersion can be as high as a few thousand km/s in the dense central regions of Virgo, where the gravitational potential is the highest. Model-corrected distances are thus uncertain in the proximity of Virgo and in particular at the caustics. This was illustrated also in Fig.~\ref{fig:cfr_dist}: the scatter between model corrected and redshift independent distances indeed increases at the distance of Virgo. This results in larger uncertainties for the local densities of Virgo members with respect to those estimated for galaxies in less dense environments. To account for a possible bias, we  consider the extreme scenario where all cluster members are located at the same distance.  This yields 3D local densities for Virgo cluster galaxies that are on average $\sim$0.4~dex higher. By collapsing the line-of-sight depth of the Virgo cluster into one distance, the associated 3D densities represent an upper limit.  We discuss the implications of this further in the next sections when referring to local densities for Virgo members.

\section{Comparing the different parameterizations of environment}\label{sec:compare_envs}
\begin{figure*}
    \centering
    \includegraphics[scale = 0.36]{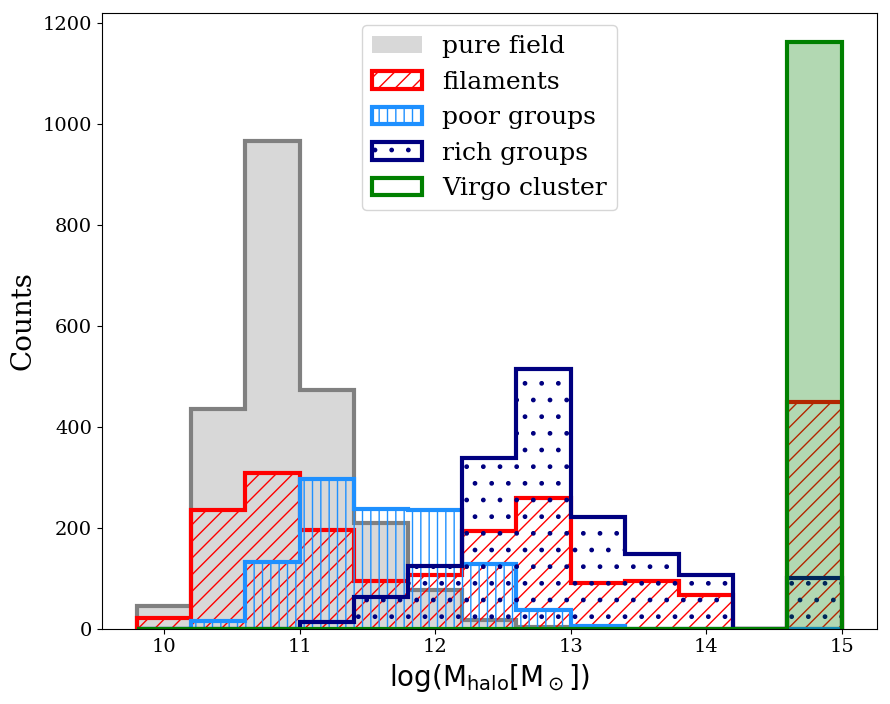}
    \includegraphics[scale = 0.37]{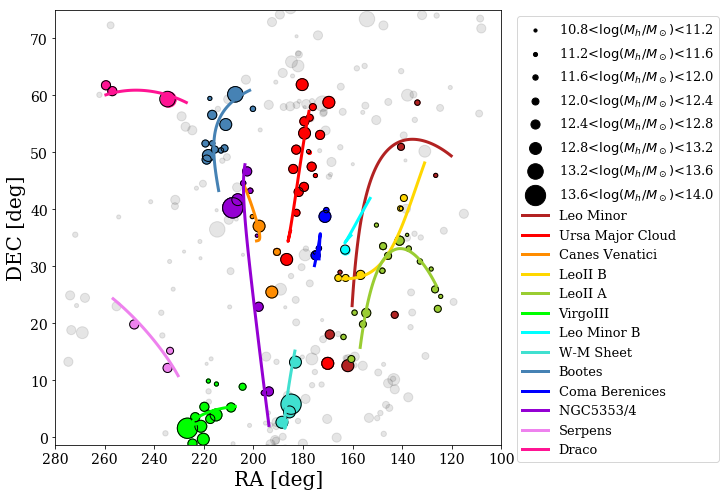}
    \caption{Left: Halo mass distribution for galaxies in the different environments, using the group memberships and halo mass distributions from     \citet{Kourkchi2017}. Right: Spatial distribution of groups and filaments around the Virgo galaxy cluster. Shaded grey points represent all groups in the velocity range 500$<v_r<3300$ km/s according to \cite{Kourkchi2017}. Lines represent the filament spines;  colored points represent the groups that share galaxies with the corresponding filament, plotted with the same color. The size of the points scales as the halo mass.\label{fig:halos}}
\end{figure*}

\begin{figure}
    \centering
    \includegraphics[scale=0.5]{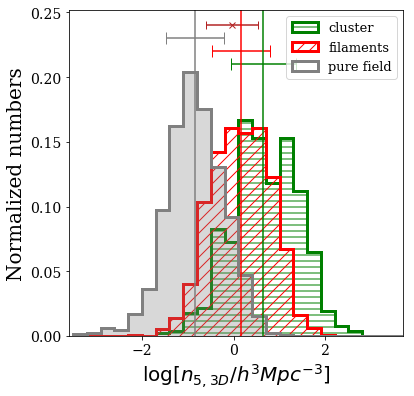}
    \caption{Volume 3D number density distribution for galaxies in the filaments (red), in the  Virgo cluster (green) and in the field (grey), for galaxies above the absolute magnitude completeness limit. The reported errors are standard deviations, while the error of the means are much smaller, of the order of 0.05~dex at most.   The firebrick cross and errorbar show the median density for filaments when the W-M Sheet and the Ursa Major Cloud are removed from the filament sample. 
    \label{fig:density_distr}}
\end{figure}

We are now in position to compare the different metrics adopted to define the environment: the cluster, filament, and group memberships,  
local densities and the halo masses of the hosting structure.  
By looking for possible differences between the different environments, we can gain insights on the physical mechanisms acting at the different scales. 

Figure \ref{fig:halos} focuses on the global environment: the left panel shows the halo mass distribution of the different subsamples. A correlation between halo mass and environment appears clear: the different environments span different ranges in halo masses, and the typical halo mass increases from pure field galaxies - peaking around $M_{halo} = 10^{11} M_\odot$ - to  poor groups to rich groups to the cluster.
The separation in halo mass between poor and rich groups is quite evident and occurs at  $M_{halo} \sim 10^{12.2} M_\odot$.
This rather clear cut justifies our choice to use the richness of 5 members as threshold to separate poor and rich groups.

Turning the attention to  filaments, which are the focus of our analysis, we observe that they span a  wide range in halo mass and the distribution is rather flat, suggesting that filaments can also host or, more generally, be linked to structures of different halo masses. About $\sim400$ filament galaxies are also formally associated with the Virgo cluster halo itself. This is because, as already mentioned, the Ursa Major cloud and the W-M sheet extend up to the Virgo cluster region itself. 

To further investigate the connection between filaments and groups, the right panel of  Fig. \ref{fig:halos} shows the position of the groups identified by \cite{Kourkchi2017}  overplotted with the position of the filaments, identified by their spines for the sake of clarity. Some filaments do not to contain any rich groups, while others clearly include groups, with varying incidence (from few to 50\% of the galaxies). 
In particular, the NGC3535/4 filament is named for the rich group where the filament seems to terminate, i.e., the filament knot \citep{Kim2016}. The VirgoIII filament is an alignment of several groups (e.g., NGC 5248, 5364, 5506, 5566, 5678, 5746, and 5775) and terminates to the East with the NGC~5846 group.\footnote{\url{http://www.atlasoftheuniverse.com/galgrps/viriii.html}}

When investigating galaxy properties in filaments, it is therefore important to consider the presence or absence of galaxy groups. 
We note that the spine of the Ursa Major Cloud seems very short when compared to the distribution of member groups presented in Fig.\ref{fig:positions_fil}. This is merely a projection effect, as the closest point of the Ursa Major cloud to Earth is only 2.6 $h^{-1}$ Mpc. At this distance, filament member galaxies, defined as those within 2 $h^{-1}$ Mpc from the spine, are spread over 30 degrees on the plane of the sky and appear to have a large projected distance from the southern end of the spine.

Next, we correlate the global and local environments by investigating the local density distribution in galaxies in different global environments (Fig.~\ref{fig:density_distr}). 
Cluster, filament, and pure field galaxies  cover different density ranges, with pure field galaxies lying preferentially at lower densities, and cluster galaxies at the highest ones. Filament galaxies span an intermediate  range of local densities. This agrees with predictions from simulations \citep{Cautun2014} and with what we already showed in \citetalias{Castignani2021}, though for a smaller sample of filament galaxies.  Nonetheless,  there is non-negligible overlap among the different distributions, indicating that there are low density regions in the cluster and relatively dense regions in the field. The median density of the filament galaxies is considerably influenced by the Ursa Major Cloud and the W-M Sheet, which host galaxies simultaneously belonging to both the clusters and the aforementioned structures, at the high density tail of the distribution. 

\begin{figure}
    \centering
    \includegraphics[scale=0.6]{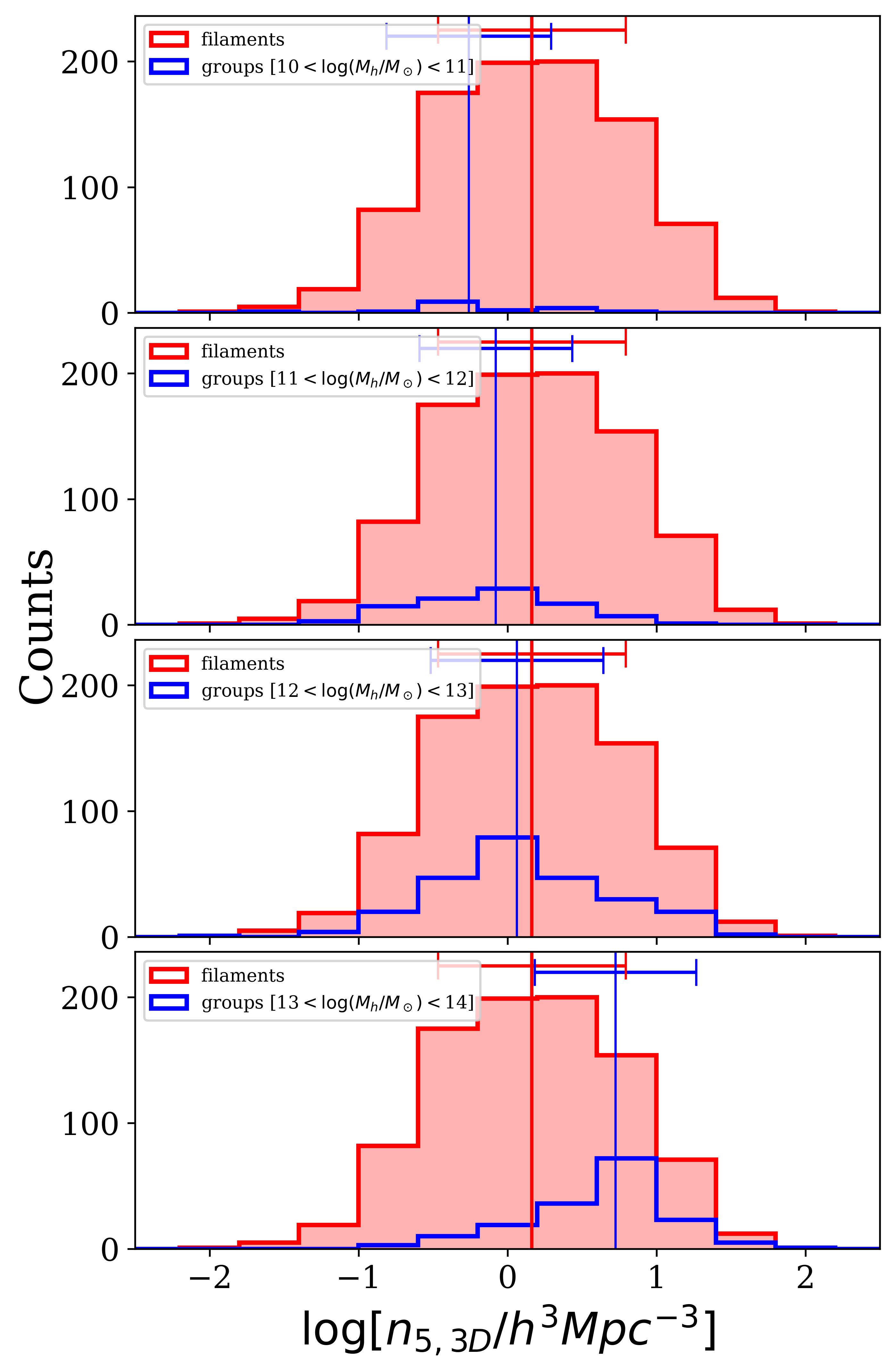}    \caption{Volume number density distribution for galaxies in groups of different halo masses and in filaments (blue histograms). Red histograms show the overall distribution for filament galaxies. The solid vertical lines show the median values of the distributions, while the error bars show the $1\sigma$ uncertainties.\label{fig:distribution_density_groups}}
\end{figure}

The Ursa Major Cloud and the W-M Sheet are not the only structures sharing galaxies with other systems: as already mentioned, other filaments share galaxies with groups of different richness. It could therefore be possible that the large density range probed by filaments is driven by the presence/absence of groups. In Figure \ref{fig:distribution_density_groups} we therefore compare the density distribution of filament galaxies (red histograms) to the density distribution of group galaxies (blue histograms) that are also in the filaments, subdivided in bins of halo mass.  A shift towards larger densities when increasing the halo mass is clearly visible, confirming that filament galaxies at the highest densities are likely also members of a group.

\begin{figure*}
    \centering
    \includegraphics[width=1.0\textwidth]{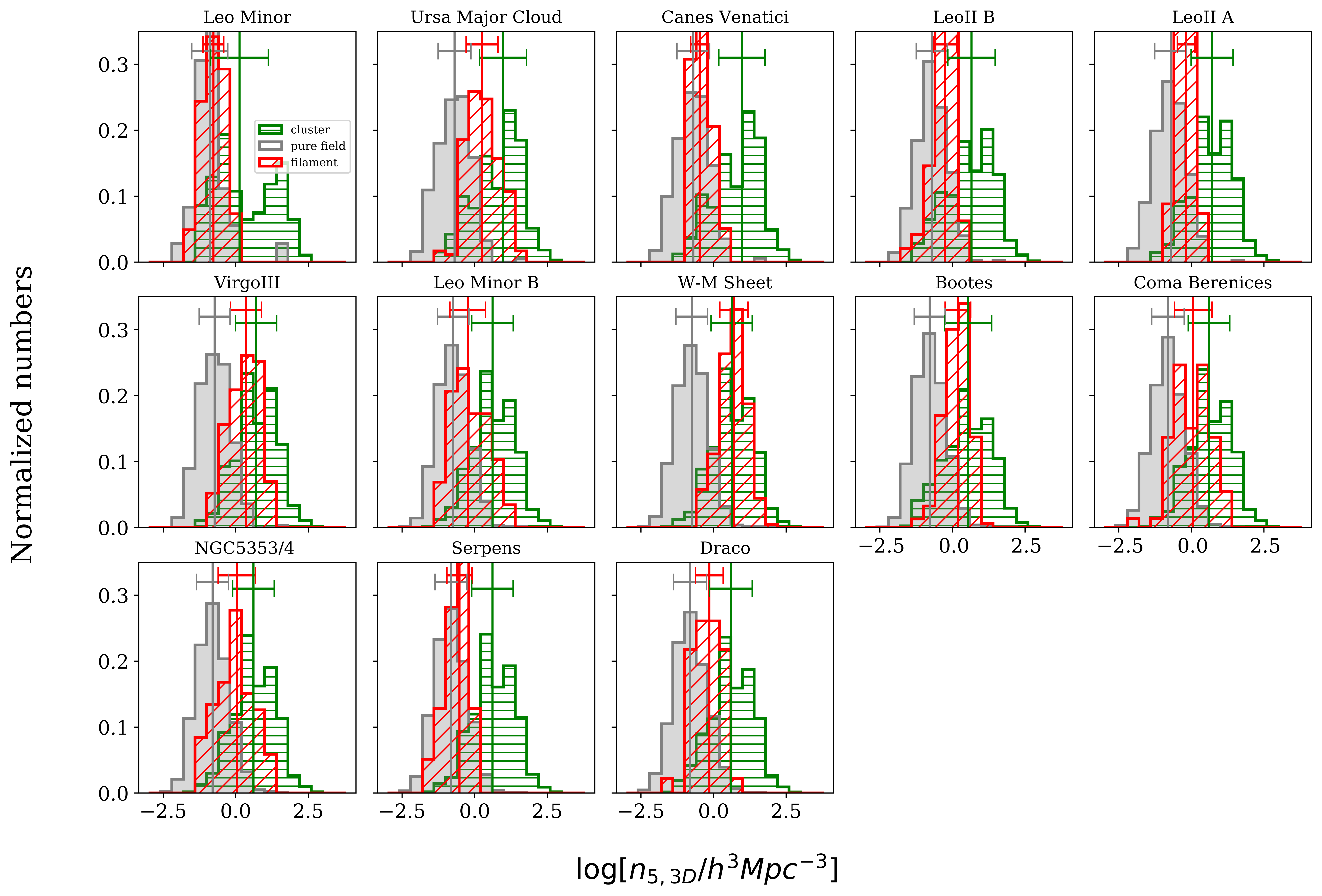}
    \caption{Volume number density ($n_{5,3D}$) distribution for galaxies in each filament separately (red). For each filament, the proper absolute magnitude limit (see Table~\ref{tab:filaments_mag}) has been adopted, to increase the statistics. For comparison, also the distributions of galaxies in the  Virgo cluster (green) and in the field (grey) are also reported, above the same completeness limit and limiting the sample to the same velocity. 
    Vertical lines represent median values, horizontal lines the standard deviation, representing the scatter of the distribution.\label{fig:distribution_density_fil}}
\end{figure*}

Finally, we inspect the density distribution of the different filaments, separately, to determine if overall all filaments behave similarly or if there is a wide  filament to filament variation. To increase the statistics, for each filament we use its proper completeness limit (see Table~\ref{tab:filaments_mag}), and extract from the field and cluster samples only galaxies above the same limit and located up to the same distance. 
Figure~\ref{fig:distribution_density_fil}  highlights that different filaments are characterized by different density distributions, taking into account both the median and the range in density. 

To conclude, the main result of this section is that even though the local and global parameterizations of the environment agree qualitatively with each other, there is no clear one-to-one correlation between the two. This demonstrates that contrasting the variation of galaxy properties as a function of the global and local environment separately is important in identifying the acting physical mechanisms.

\section{Properties of the galaxies in the different environments}\label{sec:prop_galaxies}

In this section we provide an overview of the properties of galaxies located in the different environments.
We  consider the  de Vaucouleurs morphological parameter (simply called morphology from now on, Sect.~\ref{sec:morphology})  and the presence of bars (Sect.~\ref{sec:bars}).
These parameters are taken from the HyperLeda catalog. Above the completeness magnitude limit $M_r=-15.7$, 3485/3530 galaxies have a value of morphology, and 3450/3530 have information on the presence or absence of a bar. 

The HyperLeda catalog also provides information on the position angle of each galaxy. Similarly to \citetalias{Castignani2021}, we  measure the projected orientation $\theta_{\rm alignment}$ between the major axis of each filament galaxy and the direction of the filament spine, estimated at the point of minimum distance from the galaxy. The alignment is thus the galaxy position angle, 0~deg~$\leq\theta_{\rm alignment}\leq$~90~deg, with respect to the projected orientation of the filament in the plane of the sky.
In Sect.~\ref{sec:alignments} we search for possible features in the alignments of galaxies in the filaments.

\subsection{Morphologies}\label{sec:morphology}
We investigate the morphological properties of galaxies as a function of their global environments (cluster, filaments, groups, field) and the associated local parameterization, in terms of local densities. We will distinguish galaxies between early type (de Vaucouleurs morphological type $T<$0, ET) and late type ($T\geq$0, LT). 

\begin{figure*}
    \centering
    \includegraphics[scale=0.6]{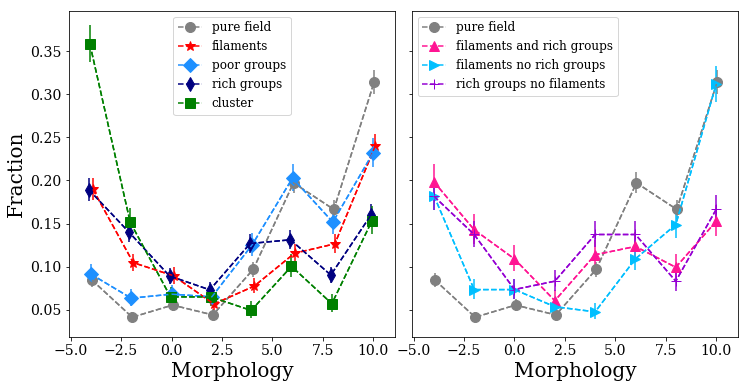}
    \caption{Fraction of galaxies of different morphological type in the different environments, as described in the legend.  A small arbitrary horizontal shift has been applied to the points for the sake of clarity.\label{fig:morph}}\end{figure*}

\subsubsection{Morphology and the environment}
Figure~\ref{fig:morph} shows the incidence of each morphological type  in the different global environments. As seen in the left panel, there is a clear dichotomy in the morphology of cluster and pure field galaxies, which have preferentially early- and late-type morphology, respectively. 

Galaxies in poor groups follow quite closely the trend of the pure field galaxies, while overall rich groups and filaments have intermediate behaviors, with an excess of ET galaxies  with respect to the pure field, and an excess of late type galaxies with respect to the cluster.

To understand if the trends in filaments depend on the presence/absence of massive groups within them or if filaments are truly a site of transformations, in the
right panel of Fig.~\ref{fig:morph} we compare the morphological distribution of galaxies only belonging to filaments to those belonging simultaneously to a filament and a rich group. Filament galaxies that are not in rich groups exhibit a bimodal morphological distribution: galaxies have either a very early or late type morphology, while intermediate values are less favoured. The observed  excess of early type galaxies with respect to the pure field suggests that filaments induce a morphological transformation,  even when groups within them are not included.

In contrast, galaxies of rich groups, either in filaments or not, show a fairly uniform distribution in morphological type, suggesting that rich groups act as the main driver for the suppression of the LT galaxy excess that is typical of the pure field. The fraction of the earliest type is the highest when galaxies are both in rich groups and filaments, suggesting that the combination of the two environments promotes transformations. When considering poor groups, we verified that differences between galaxies in both filaments and groups and only in filaments disappear, indicating that poor groups do not play a major role in inducing morphological transformations.

\begin{figure}
    \centering
    \includegraphics[scale=0.45]{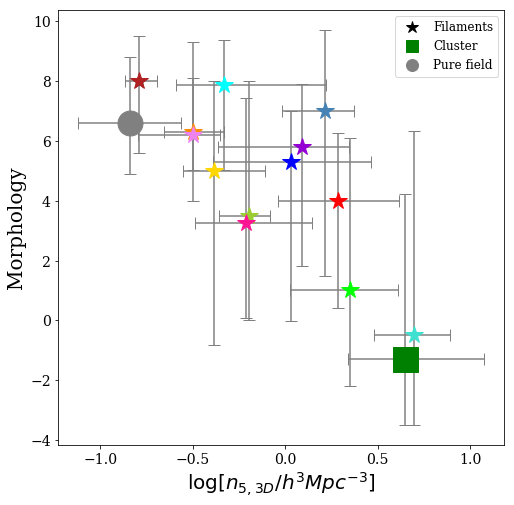}
    \caption{Median morphological parameter for filaments (coloured stars), pure field (grey point) and the Virgo cluster (green square). Error bars represent 1$\sigma$ dispersion. The color code for filaments is the same as in Fig.~\ref{fig:positions_fil}.  \label{fig:morph_ld_fil}}
\end{figure}

The results above highlight that the dependence of the morphology on the global environment is complex. This is particularly true for filaments, which span four orders of magnitude in local density. We therefore look for any morphological trends as a function of local density.
Figure~\ref{fig:morph_ld_fil} shows the median morphological  $T$-type plotted against the median local density for each filament, separately. Cluster and pure field values are shown for comparison. Overall, even though the scatter is large, the two quantities are anti-correlated: denser structures tend to be dominated by early-type galaxies. Filaments are intermediate between the pure field and the cluster, and a large filament to filament variation is detected on both axes. A few structures, i.e., Leo Minor, Canes Venatici, Leo Minor B, and Serpens show almost no ET galaxies. These are filaments with only a few groups (Fig.~\ref{fig:halos}, right) and with the lowest average densities. In contrast, VirgoIII, Ursa Major Cloud, and the W-M sheet have on average the highest local densities, higher fractions of early-type galaxies, and are rich in groups. We remind the reader that VirgoIII is an alignment of 
several groups, while both the W-M Sheet and the Ursa Major Cloud are connected to Virgo cluster itself. This may explain at least partially their higher local densities and the prevalence of ET galaxies.

To conclude, the above results show that the ET galaxies are largely present already in filaments, which support the scenario that morphological transformations may occur well before galaxies fall into the cluster core.

\begin{figure*}
    \centering
    \includegraphics[scale=0.35]{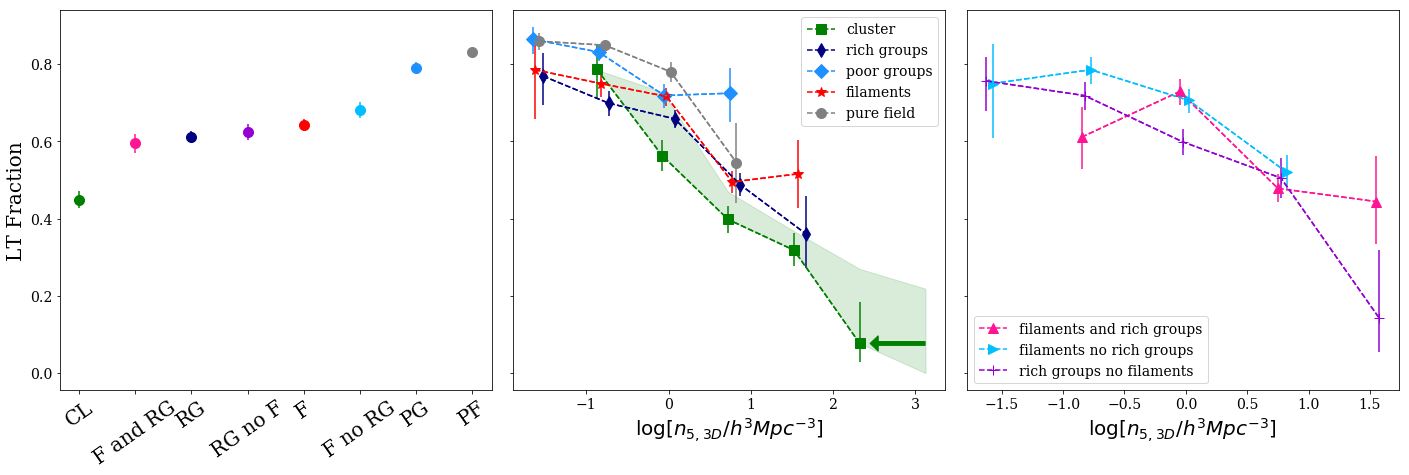}
    \caption{LT fraction as a function of different environments (left) and local density (center, right). Different environments are considered in the left panel: cluster galaxies (CL), filament galaxies (F), pure field galaxies (PF), galaxies in rich (RG) and poor (PG) groups, as well as a combination of these classes, at intermediate environments.
    In the central panel, the right border of the green dashed area defines the conservative upper limit to the local density for Virgo cluster galaxies. A small arbitrary horizontal shift has been applied to the points for the sake of clarity.\label{fig:morph_frac}}
\end{figure*}

\subsubsection{Morphological fractions in the different environments} 
We now quantify the variation of the morphological fraction with environment, more specifically in terms of the LT fraction, i.e. the number of galaxies with a late type morphology over the total.
The left panel of Fig.~\ref{fig:morph_frac} shows that, considering the different global environments, the fraction monotonically increases from the cluster ($\sim 40\%$) to the pure field ($>80\%$), while filaments have an intermediate fraction ($60\%$). 
Interestingly, galaxies that are both in groups and filaments have a lower probability of being LT than  both galaxies in groups only and sources in filaments only.  
This result again points to the scenario according to which both filaments and groups affect morphology, separately, and their effect is amplified for galaxies simultaneously in both environments. 
We verified that we obtain similar results when considering the halo mass of the hosting structure, with the fraction of LT decreasing with increasing halo mass.

We are now in the position to investigate the so-called morphology-density relation \citep{Dressler1980}. This relation was first established for clusters only, then groups \citep{Postman1984}, and we now inspect it also in  other global environments (central and right panels  in Fig.~\ref{fig:morph_frac}), to determine which environmental definition plays the major role. 
In each global environment taken separately, we see a decline of the LT fraction with increasing density. Nonetheless, the global environment does play a role in shaping the LT fraction: at a fixed density, the LT fraction increases from cluster, rich groups, filaments, poor groups, and to the pure field. As discussed in Sect.~\ref{sec:ld}, local densities in Virgo are more uncertain than for the other global environments. We have therefore computed the cluster morphology density relation using the density estimates obtained assuming that all cluster galaxies are at the exact same distance.  This provides a conservative upper limit on the local density estimates, as the distance along the line-of-sight between cluster galaxies is artificially set to zero.  This compression of distances yields local densities that are $\sim0.4$~dex higher, on average, than the actual estimate for the local density of cluster galaxies. 
The right border of the green area in Fig.~\ref{fig:morph_frac} shows the relationship derived when using the upper limits on local density. Even assuming the upper limits as true values for the local density of cluster galaxies, their associated LT fractions only tentatively reach those of filament galaxies. This result shows that uncertainties associated with the local densities of cluster galaxies do not impact our results: the observed differences between the global environments considered remain.

In the right panel of Fig.~\ref{fig:morph_frac} we  look for other  possible differences when considering filament and rich group galaxies, in all possible combinations. While these environments showed different  morphology distributions (Fig.~\ref{fig:morph} right), these differences disappear in the LT fraction vs. density plot. This suggests that the overall density - morphology relation is similar for groups and filaments, even if there are measurable differences in the morphological composition of their galaxy populations.

Finally, we investigate the dependence of LT fraction as a function of distance to the cluster and to the filament spines (plots not shown). 
The LT fraction of filament, field, and group galaxies is flat, up to the largest cluster-centric distances ($\sim30~h^{-1}$~Mpc). A similarly flat behaviour is observed as a function of the distance to the filament  spines for filament members.  In particular, to appreciate a trend (if any) we should reach larger distances from the filament spine than 2~$h^{-1}$~Mpc, i.e., the radius up to which filament membership are assigned. This is a consequence of the fact that at larger distances we have the strongest density contrast with respect to the central regions close to the filament spines.

\subsection{Bars}\label{sec:bars}
\begin{figure*}
    \centering
    \includegraphics[scale=0.35]{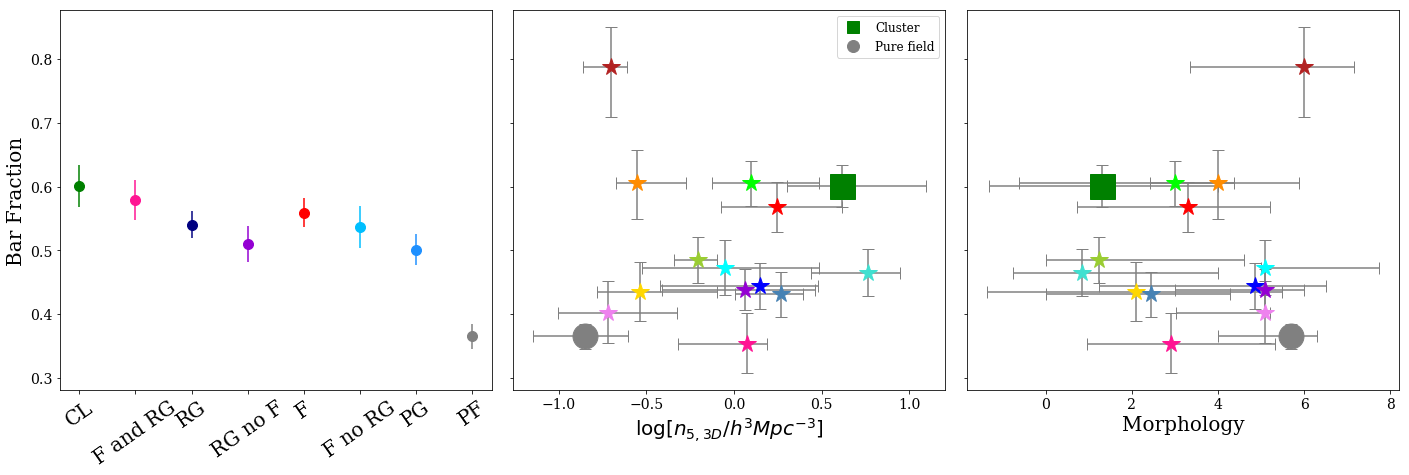}
    \caption{Bar fraction as a function of the environment (left), local density  (center), and morphology (right). Only lenticular galaxies and normal spirals are considered, see text for details. For the left panel the different environments are reported as in Fig.~\ref{fig:morph_frac} (left). In the central and right panels filaments (coloured stars), pure field (grey point), and the Virgo cluster (green square) are distinguished. The color code for the filaments is the same as in Fig.~\ref{fig:positions_fil}.\label{fig:bars}}
\end{figure*}

We now investigate the presence of bars in our sample. In this subsection only, we conservatively exclude both elliptical and irregular galaxies because they typically do not show evidence of bars. We thus limit ourselves to lenticular and spiral galaxies (i.e., $-3\leq T\leq8$), and we consider as barred galaxies those sources that are classified as barred by HyperLeda. 

In Fig.~\ref{fig:bars} we investigate the bar fraction as a function of the global environment (left panel) and as a function of the local density (center), and morphology (right), considering each filament separately. There is a light, but rather systematic decrease in the bar fraction from the cluster, to filaments, to the pure field. This trend might be at least  partially due to local density. As illustrated in the central panel, a mild trend towards a higher bar fraction for increasing local density is observed. Filaments are intermediate between the cluster and the pure field, showing bar fractions of $\sim0.35$ and $\sim0.6$  for the field and cluster, respectively. In contrast, as shown in the right panel of Figure~\ref{fig:bars}, we do not observe any clear trend of the bar fraction as a function of the average/median morphology. 

In \citetalias{Castignani2021} we showed that the fraction of galaxies with star formation below the main sequence monotonically increases in filaments with increasing local density. The observed trend for the bar fraction as a function of local density could thus be related to the fact that the presence of bars may favor the cessation (quenching) of star formation, as suggested by a number of studies \citep[e.g.,][]{JamesPercival2016, Fraser2020,  Newnham2020}.

A large scatter is nonetheless 
observed when comparing the different filaments, with the nearby ones  preferentially showing the highest bar fractions. An example is the nearby Leo Minor filament, that has a very high bar fraction $\sim0.8$ and low average density, while its galaxy population is mostly composed by LT galaxies. Note however, that these results are based on only 7 galaxies, while the number of barred galaxies in the other filaments range from 15 to 82.
In addition, we note that the bar identification is a very delicate task and that the bar detection in HyperLeda has been attempted only for a small fraction of  galaxies, mostly those  larger than 1~arcmin of diameter, which may cause a possible bias in the determination of the bar fraction. Furthermore, the classification likely comes from optical images, while bars are better seen in the infrared \citep{Eskridge2000}. 

Overall, from our analysis we find that bars are found in $56\%$ of lenticular and spiral galaxies in filaments.
The fractions distributed as follows: $41\%$ (39/95) for lenticulars ($-3\leq T<-1$),  $59\%$ (73/124)  for ET spirals ($-1\leq T<3$), and $60\%$ (154/257) for LT  spirals ($3\leq T\leq8$). The total fraction of galaxies with strong or weak bars should be closer to 2/3 as SA, SAB, and SB galaxies are in proportion 1/3 each \citep[e.g.,][]{Eskridge2000}. It is thus likely that some of the galaxies in our sample are misclassified as non-barred, as a consequence of the observational uncertainties mentioned above.

Nevertheless, our bar fractions are fairly in agreement with those found for galaxies in the local universe, in the range $\sim(45-60)$\%
\citep{Marinova_Jogee2007, Reese2007, Barazza2008}.  In particular, \cite{Aguerri2009} considered the redshift range $0.01<z<0.04$ and found fractions equal to 29\%, 55\% and 54\% for lenticulars, ET and LT spirals, respectively.  To derive these fractions the authors analyzed  the r-band images of a large sample of galaxies in SDSS down to an absolute magnitude limit of $M_r$=-20, a magnitude limit that is brighter than what we use in this work. By using the same magnitude cut adopted by the authors we obtain even higher fractions for all considered classes, with an overall bar fraction of 71$\%$. These differences highlight the difficulty in assessing an absolute bar fraction that is independent of the sample selection, the images used, and the method adopted to detect the bars.  

\begin{figure*}
    \centering
    \includegraphics[scale=0.55]{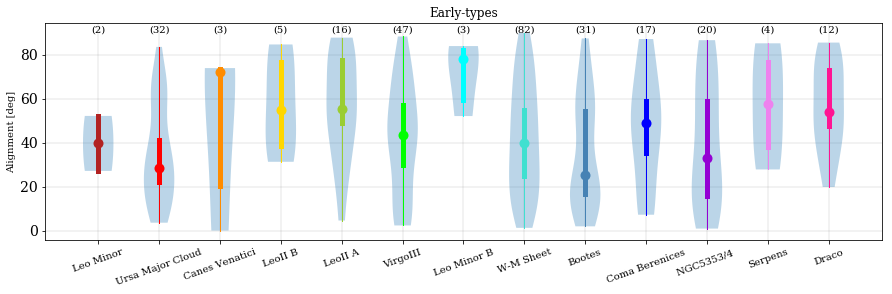}
    \includegraphics[scale=0.55]{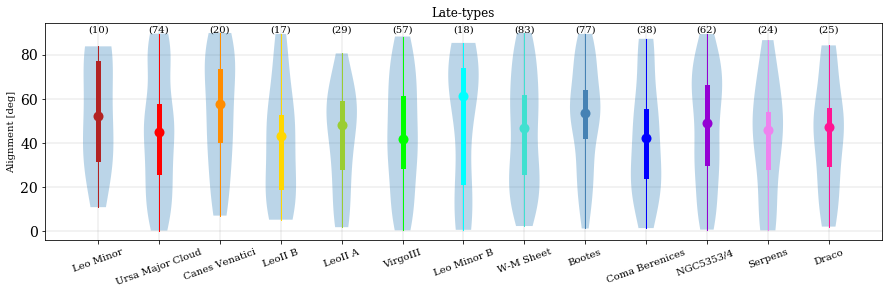}
    \caption{Violin plots of the alignments $\rm \theta_{alignment}$ for ET (top) and LT (bottom) galaxies in filaments. For each filament, medians and the interquartile ranges are also shown with circles and thick bars, respectively.  We report in parentheses the number of galaxies in each filament.\label{fig:align_distr} }
\end{figure*}

\subsection{Galaxy alignments with respect to the filament spines}\label{sec:alignments}
We conclude the overview of galaxy properties by investigating the alignment of filament galaxies with respect to the filament spine.  

Overall, for each of the filaments we verified that the distribution of $\theta_{\rm alignment}$ is fairly uniform, with mean alignments around 45~deg. Previous studies \citep{Tempel2013, TempelLibes2013, Hirv2017, Codis2018,  Chen2019, Welker2020, Kraljic2021} found that the spin axis of ET and LT galaxies is preferentially perpendicular and parallel, respectively, to the filaments. The expected difference can be ultimately related to the different assembly history of ET and LT galaxies. ET galaxies are thought to be predominately formed via major mergers. During these events, the rotation axis  of  the  resulting  galaxy  tends  to  be  perpendicular  to  the merger  direction. 
For LT galaxies the assembly primarily occurs via the winding of flows, and the alignment of angular momentum with the filament spine is related to the regions outside filaments, namely sheets, where most of the gas is falling in from \citep{TempelLibes2013}. 

In  Fig.~\ref{fig:align_distr} we therefore inspect the alignment distributions for LT and ET galaxies, separately. 
Data distributions are shown in terms of violin plots, which give the probability density of the data at different values, smoothed by a kernel density estimator. Unlike bar graphs with means and error bars, violin plots show the distribution of all data points. The shape of the violin displays frequencies of values: the thicker part of the violin shape means that the values in that y-axis section of the violin have higher frequency, and the thinner part implies lower frequency. Violin plots also highlight the maximum extension of the data, and the presence of different peaks, their position and relative amplitude. The maximum width of each violin is set the same for all galaxies, for display purposes.

While for LT galaxies the average alignments scatter around 45~deg for all filaments, for ET galaxies we do see a higher filament by filament variation, with median $\theta_{\rm alignment}$ values ranging from $\sim20$ to 80 degree.  This large scatter may be due to the limited number of ET galaxies in each filament, which is reported in parentheses above each violin in the Figure.

We did not find any statistically significant difference when comparing the overall distribution of $\theta_{\rm alignment}$ of ET and LT galaxies with the Kolmogorov-Smirnov test. This is partially at odds with the aforementioned  studies; 
however, the absence of a significant difference  may be due to uncertainties associated with the determination of the alignment angle, as the analysis is done in projection and relies on the position angles of the galaxy and the filament, estimated locally, which are both uncertain. Similarly, no difference 
has been found in $\theta_{\rm alignment}$, when considering barred and non-barred galaxies, separately. No trend of $\theta_{\rm alignment}$ as a function of distance to the filament spine has been found.

\section{Summary and Conclusions}\label{sec:conclusions}

We have presented a comprehensive catalog of galaxies extending up to $\sim12$ virial radii in projection from the Virgo cluster, with the intent of characterizing the complex network of filamentary structures around Virgo and investigating the role of filaments in galaxy evolution.  We select  spectroscopically-confirmed galaxies from HyperLeda, the NASA Sloan Atlas, NED, and ALFALFA, assembling a sample of galaxies in the region 
$ 100^\circ < RA < 280^\circ$, $-1.3^\circ < DEC < 75^\circ$, with recession velocities in the range $500 < v_r < 3300$~km/s. These cuts ensure that both Virgo and its main filaments in the Northern hemisphere are included. The final catalog contains { 6780} galaxies, { 3528} of which are brighter than the absolute magnitude limit $M_r=-15.7$ \citep[$\simeq M^\star_r+3$,][]{Blanton2005b}.  

To characterize the environment around Virgo, we adopt a  number of parameterizations that trace different scales.
By exploiting a tomographic approach, we recover 13 filaments, spanning several Mpc in length.  

We then assign filament memberships relying on the 3D distance of the galaxies from the filament spines, which we release for all 13 considered filamentary structures.
We also identify the cluster members both in the 3D Super Galactic coordinate frame and also considering the cluster region in phase space.

To further characterize the environments of our catalog galaxies, we match our sample to  \cite{Kourkchi2017}'s group catalog, to select galaxies in groups and extract for each galaxy of the sample the halo mass estimate of the hosting structure. Finally, we quantify the local environment using surface (2D) and volume (3D) local densities in terms of 5th-nearest neighbors. We make available the catalogs of galaxies and of the aforementioned environments.

We then characterize galaxy morphology and spin alignment of galaxies in filaments and discuss the different parameterizations of environment. 
The main results of our analysis are:
\begin{itemize}
    \item     By fitting an exponential model to the  distribution of galaxies, averaged in cylindrical shells around each filament spine, we find that long $>17h^{-1}$~Mpc filaments have low characteristic radii $r_0<1~h^{-1}$~Mpc (along the direction perpendicularly to the filament spine) and the lowest density contrasts with respect to the field. Shorter filaments have a larger range of values of both the density contrast and characteristic radius, and extend to higher values in each.
    \item Filament galaxies span a wide range of $\sim~4$~dex in both local density and halo mass of the hosting structure (e.g., group). Values range at the low end from those typical of the field to values found in the Virgo cluster at the high end. The high dispersion found for the filaments is ultimately due to the large filament to filament variation and to the fact that some filaments are very rich in groups, while other are poorer.
    \item A decline of the late-type (LT) fraction with increasing local density is observed in all considered global environments (field, filaments, groups, and cluster). At fixed local density, filaments appear to be an intermediate environment between the field and the cluster, with a decline resembling that of rich groups. The local density alone is thus not sufficient to explain the dependence of the LT fraction with the Mpc-scale environment.  
    \item The average fraction of barred galaxies decreases from the highest density regions of the cluster, to the field at the lowest density.
    Filaments show an intermediate and broad range in the fraction of barred galaxies, with a large filament to filament variation, which reflects the large dispersion for filament galaxies observed also in local density and morphology.
\item We find no clear dependence of the projected orientation of the galaxy major axis with the filament spine for either early or late type galaxies. Similarly, we did not find any clear trend for the considered properties of filament galaxies as a function of their distance to the spines. However, it is important to note that we only consider filament members to be those galaxies closer than $2~h^{-1}$~Mpc from the filament spine.  While this radius allows us to minimize contamination from field galaxies, it does make it hard to assess whether trends would exist if we included galaxies at larger distances.

\end{itemize}

 \acknowledgments{
 
 The authors thank the hospitality of International Space Science Institute (ISSI) in Bern (Switzerland) and of the Lorentz Center in Leiden (Netherlands). Regular group meetings in these institutes allowed the authors to make substantial progress on the project and finalize the present work. 
 
 GC acknowledges financial support from the Swiss National Science Foundation (SNSF).  BV acknowledges financial contribution  from the grant PRIN MIUR 2017 n.20173ML3WW\_001 (PI Cimatti) and from the INAF main-stream funding programme (PI Vulcani). RAF gratefully acknowledges support from NSF grants AST-0847430 and AST-1716657. GHR acknowledges support from NSF-AST 1716690.  
 
 This research has made use of the NASA/IPAC Extragalactic Database (NED) which is operated by the Jet Propulsion Laboratory, California Institute of Technology, under contract with the National Aeronautics and Space Administration. We acknowledge the usage of the HyperLeda database.\footnote{http://leda.univ-lyon1.fr}  This research made use of Astropy\footnote{http://www.astropy.org},
a community developed core Python package for Astronomy \citep{astropy2013,astropy2018}, matplotlib \citep{Hunter2007}, and TOPCAT \citep{Taylor2005}.}

\appendix

\section{Catalogs}\label{sec:catalogs}

With this paper, we release a number of catalogs: 
the main galaxy catalog, the catalog of the environmental properties, and the catalog with the filament spines.
The main galaxy catalog is shown in Table~\ref{tab:main} for a subsample of 10 galaxies. The table is presented in its entirety in the online version of the article. The columns indicate:

\begin{itemize}
    \item Column (1) --- VFID, a unique serial number, with galaxies sorted by declination from north to south; 
    \item Columns (2) and (3) --- right ascension and declination at epoch J2000 (in degrees);  
    \item Column (4) --- $v_{r}$ heliocentric velocity (units of km~s$^{-1}$); 
    \item Column (5) --- $V_{cosmic}$ cosmic recession velocity (units of km~s$^{-1}$) obtained from a redshift-independent distance from \cite{Steer2017} when available or from $V_{model}$ as described in Sect.~\ref{sec:dist}; 
    \item Column (6) --- $V_{model}$ model recession velocity (units of km~s$^{-1}$) obtained from \cite{Mould2000} model, as described in Sect.~\ref{sec:dist}; 
    \item Column (7) --- HyperLeda name;
    \item Column (8) --- NED name; 
    \item Column (9) --- PGC ID; 
    \item Column (10) --- NSAID from the v0 catalog; 
    \item Column (11) --- NSAID from the v1 catalog; 
    \item Column (12) --- {\em Arecibo Galaxy Catalog} (AGC) name; 
    \item Column (13) --- boolean flag, where True indicates that the galaxy has a CO observation from \citetalias[][]{Castignani2021};
    \item Column (14) --- boolean flag, where True indicates that the galaxy is in the ALFALFA $\alpha.100$ catalog \citep{Haynes2018}.
\end{itemize}

\begin{sidewaystable*}
\begin{center}
\scriptsize
\setlength\tabcolsep{3.0pt} 
\tablenum{7} 
\caption{Main Catalog with Cross IDs\label{tab:main}  } 
\begin{tabular}{|c|c|c|c|c|c|c|c|c|c|c|c|c|c|}
\hline 
\toprule 
VFID   & RA &	DEC &	$v_{r}$ & $v_{\rm cosmic}$ &  $v_{\rm model}$  & HL~name & NED Name & PGC & NSA V0 & NSA V1 & AGC  & CO  & A100  \\ 
& (deg, J2000) & (deg, J2000) & $\rm km~s^{-1}$ & $\rm km~s^{-1}$ & $\rm km~s^{-1}$ & & & &  & & && \\ 
(1) & (2) & (3) & (4) & (5) & (6) & (7) & (8) & (9) & (10) & (11) & (12)& (13) & (14) \\ 
\hline 
\hline 
3000 & 165.930807 & 28.88713 & 708 & 1189 & 644 &NGC3510& NGC 3510& 33408 & 100677& 472983& 6126&False&True\\ 
3001 & 149.190405 & 28.82596 & 510 & 687 & 495 &UGC05340& UGC 05340& 28714 & 136251& 623560& 5340&False&True\\ 
3002 & 194.776643 & 28.81178 & 998 & 1177 & 1177 &PGC1846725& WISEA J125906.48+284842.6& 1846725 & 89104& 427578& -999&False&False\\ 
3003 & 157.778262 & 28.79659 & 1425 & 1732 & 1732 &NGC3265& NGC 3265& 31029 & 107764& 497691& 5705&True&True\\ 
3004 & 129.582068 & 28.78993 & 2669 & 2842 & 2842 &PGC3095094& WISEA J083819.66+284723.7& 3095094 & 135383& 622813& -999&False&False\\ 
3005 & 181.303258 & 28.78191 & 3153 & 3395 & 3395 &UGC07072& UGC 07072& 38268 & 102495& 478264& 7072&False&True\\ 
3006 & 250.089288 & 28.76552 & 976 & 1291 & 1291 &SDSSJ164021.43+284555.9& SDSS J164021.43+284555.9& 4123676 & 69715& 343115& 262737&False&True\\ 
3007 & 225.279839 & 28.76086 & 1821 & 2158 & 2158 &SDSSJ150107.16+284539.2& WISEA J150107.08+284539.7& 4443809 & -999& -999& 733373&False&True\\ 
3008 & 142.246120 & 28.75796 & 1228 & 1459 & 1459 &PGC1845056& SDSS J092859.06+284528.5& 1845056 & 84921& 410169& 194058&False&True\\ 
3009 & 128.807646 & 28.75335 & 2052 & 3525 & 2244 &UGC04482& UGC 04482& 24104 & 156791& 647654& 4482&False&True\\ 
\hline 
\hline 
\end{tabular} 
\end{center} 
\tablecomments{This table is published in its entirety in machine-readable format.  A portion is shown here for guidance regarding its form and content.  Galaxies without a corresponding ID in columns $9-12$ are denoted as $-999$.}\end{sidewaystable*} 

\begin{sidewaystable*}
\begin{center}
\scriptsize
\setlength\tabcolsep{3.0pt} 
\tablenum{8} 
\caption{Environmental Properties of Catalog Galaxies\label{tab:environment}  } 
\begin{tabular}{|c|c|c|c|c|c|c|c|c|c|c|c|c|c|c|}
\hline 
\toprule 
VFID &SGX &SGY &SGZ &n$_{5,2D}$ &err(n$_{5,2D}$) &n$_{5,3D}$ &err(n$_{5,3D}$) &Nearest Filament &$\rm D_{Filament}~2D$ &$\rm D_{Filament}~3D$ &Filament Memb &Group &Cluster &Pure Field \\ 
 &$h^{-1}$~Mpc &$h^{-1}$~Mpc &$h^{-1}$~Mpc &$h^{2}$~Mpc$^{-2}$ &$h^{2}$~Mpc$^{-2}$ &$h^{3}$~Mpc$^{-3}$ &$h^{3}$~Mpc$^{-3}$ & &$h^{-1}$~Mpc &$h^{-1}$~Mpc & & & & \\ 
(1) &(2) &(3) &(4) &(5) &(6) &(7) &(8) &(9) &(10) &(11) &(12) &(13) &(14) &(15) \\ 
\hline 
\hline 
VFID3000 &2.0 & 11.3 & -3.1 & 1.9 & 0.9 & 0.2 & 0.1 & Coma\_Berenices &1.2 & 2.0 & 0 &2 &0 &0 \\ 
VFID3001 &2.0 & 5.8 & -3.2 & 0.4 & 0.2 & 0.1 & 0.0 & Leo\_Minor &1.4 & 1.5 & 1 &0 &0 &0 \\ 
VFID3002 &0.3 & 11.6 & 1.7 & 0.9 & 0.4 & 0.2 & 0.1 & Canes\_Venatici &1.4 & 1.4 & 1 &0 &0 &0 \\ 
VFID3003 &3.9 & 15.7 & -6.3 & 4.5 & 2.0 & 0.5 & 0.2 & LeoII\_B &0.1 & 1.9 & 1 &2 &0 &0 \\ 
VFID3004 &12.4 & 18.3 & -17.9 & 0.4 & 0.2 & 0.1 & 0.0 & LeoII\_A &4.8 & 7.4 & 0 &0 &0 &1 \\ 
VFID3005 &2.7 & 33.8 & -1.8 & 2.7 & 1.2 & 0.7 & 0.3 & Coma\_Berenices &4.1 & 4.1 & 0 &0 &0 &1 \\ 
VFID3006 &0.3 & 7.1 & 10.7 & 0.1 & 0.0 & 0.0 & 0.0 & Serpens &5.5 & 6.8 & 0 &1 &0 &0 \\ 
VFID3007 &-0.6 & 17.7 & 12.3 & 1.1 & 0.5 & 0.1 & 0.0 & Serpens &5.2 & 6.4 & 0 &0 &0 &1 \\ 
VFID3008 &4.9 & 11.3 & -7.7 & 0.4 & 0.2 & 0.0 & 0.0 & LeoII\_B &0.4 & 2.5 & 0 &0 &0 &1 \\ 
VFID3009 &15.6 & 22.4 & -22.4 & 0.1 & 0.0 & 0.0 & 0.0 & LeoII\_A &10.3 & 14.2 & 0 &1 &0 &0 \\ 
\hline 
\hline 
\end{tabular} 
\end{center} 
\tablecomments{This table is published in its entirety in machine-readable format.  A portion is shown here for guidance regarding its form and content.}\end{sidewaystable*}

Galaxy environmental properties are listed  in Table \ref{tab:environment} for a subsample of 10 galaxies, while  the table for the total sample is given in the online version of the article. The columns indicate:
\begin{itemize}
   
    \item Column (1) ---VFID, galaxy unique serial number;
    \item Columns (2) - (4) --- Super Galactic X, Y, and Z coordinates, computed as described in Sect.~\ref{sec:dist};
    \item Columns (5) and (6) ---  local surface number density and $1\sigma$ Poisson uncertainty computed as described in Sect.~\ref{sec:ld}; 
    \item Columns (7) and (8) ---  local volume  number density and $1\sigma$ Poisson uncertainty computed as described in Sect.~\ref{sec:ld};  
    \item Column (9) --- Name of the nearest filament;
    \item Column (10) --- 2D distance of galaxy from the nearest filament; 
    \item Column (11) --- 3D distance of galaxy from the nearest filament; 
     \item Column (12) --- filament member flag, where 1 
     indicates that the galaxy is a filament member, i.e., within $2~h^{-1}~Mpc$ from the nearest filament spine.  
    \item Column (13) --- group membership flag, according to the group definition by \cite{Kourkchi2017}: 0 means the galaxy is not a member of a group, 1 means the galaxy is a member of a poor group (2$\leq N<$5), and 2 means  the galaxy is a member of a rich group (N$\geq$5), see text for details;
    \item Column (14) --- cluster membership flag as described in Sect.~\ref{sec:cl}, where 1 indicates that the galaxy is cluster member;
    \item Column (15) --- pure field galaxy flag, obtained as described in Sect.~\ref{sec:other_envs}, where 1 indicates that the galaxy is a pure field galaxy.
\end{itemize}

\newpage
\bibliographystyle{aasjournal}
\bibliography{main_rev3_after_proof}

\end{document}